\documentclass[11pt]{article}
\title{\vspace{30mm} \bf{Notes on the Wess-Zumino-Witten-like structure\,: \\ $L_{\infty }$ triplet and NS-NS superstring field theory} \vspace{15mm} }
\author{\Large{Hiroaki Matsunaga} \vspace{5mm}}
\date{Institute of Physics, the Czech Academy of Sciences, \\ Na Slovance 2, Prague 8, Czech Republic \\ \vspace{2mm}
Yukawa Institute for Theoretical Physics, Kyoto University, \\ Kyoto 606-8502, Japan \\ \vspace{3mm} 
Email: matsunaga@fzu.cz
\vspace{10mm}}

\usepackage[dvips]{graphicx}
\usepackage{amsmath,amscd, amssymb}
\usepackage{cite}
\usepackage{mathrsfs} 
\usepackage{amsfonts}

\usepackage{upgreek}
\makeatletter
  
  \@addtoreset{equation}{section}
\makeatother

\newcommand{\ld}{ [ \hspace{-0.6mm} [ }
\newcommand{\rd}{ ] \hspace{-0.6mm} ] }
\newcommand{\Ld}{ \big[ \hspace{-1.1mm} \big[ }
\newcommand{\Rd}{ \big] \hspace{-1.1mm} \big] }

\newcommand{\la}{\big{\langle }}
\newcommand{\ra}{\big{\rangle }}
\newcommand{\no}{\nonumber\\}
\newcommand{\niu}[1]{\vspace{7pt}\noindent\underline{{\sf\hspace{4pt}#1\hspace{4pt}}}\vspace{4pt}}

\newcommand{\teta}{{\tilde{\eta }}}
\newcommand{\ts}{\tilde{s}} 
\newcommand{\bD}{\mathbf{D}}
\newcommand{\bL }{\mathbf{L}} 
\newcommand{\cA}{\mathcal{A}} 
\newcommand{\cF}{\mathcal{F}} 
\newcommand{\cG}{\mathcal{G}}
\newcommand{\cH}{\mathcal{H}}
\newcommand{\cI}{\mathcal{I}} 
\newcommand{\cP}{\mathcal{P}} 

\allowdisplaybreaks[3]

\begin{document}
\maketitle
{\vspace{-144mm}
\rightline{\tt YITP/16-144}
\vspace{144mm}}
\begin{abstract}
In the NS-NS sector of superstring field theory, there potentially exist three nilpotent generators of gauge transformations and two constraint equations: It makes the gauge algebra of type II theory somewhat complicated. 
In this paper, we show that every NS-NS actions have their WZW-like forms, and that a triplet of mutually commutative $L_{\infty }$ products completely determines the gauge structure of NS-NS superstring field theory via its WZW-like structure. 
We give detailed analysis about it and present its characteristic properties by focusing on two NS-NS actions proposed by JHEP {\bf 1701} (2017) 022 [arXiv:1512.03379] and JHEP {\bf 1408} (2014) 158 [arXiv:1403.0940]\,. 
\end{abstract}
\thispagestyle{empty} 
\clearpage 

\setcounter{page}{1}
%\tableofcontents

\section{Introduction} 

In the previous work \cite{Goto:2015pqv}, we provided analysis of algebraic framework describing gauge invariances of superstring field theories, which we call the Wess-Zumino-Witten-like structure, and showed that there exist (alternative) WZW-like actions which are off-shell equivalent to $A_{\infty }/L_{\infty }$ actions given by \cite{Erler:2014eba}. 
In this paper, we focus on the NS-NS sector and present details of analysis and its characteristic properties: 
Some implicit or missing parts and several important properties which remain unclear in \cite{Goto:2015pqv} will be clarified. 
Through these analysis, we will see that a pair of nilpotent products, which we call an $L_{\infty }$ triplet, induces WZW-like framework and thus ensures the gauge invariances of superstring field theories of \cite{Goto:2015pqv, Erler:2014eba, Berkovits:1995ab, Erler:2013xta, Berkovits:1998bt, Okawa:2004ii, Berkovits:2004xh, Matsunaga:2013mba, Matsunaga:2014wpa, Sen:2015uaa, Saroja:1992vw, Jurco:2013qra}.\footnote{Potentially, it goes for \cite{Sen:2015uaa, Saroja:1992vw, Jurco:2013qra} and other earlier proposals. See the footnote 8 in section 5 and section 6. Note that for the NS sector, its WZW-like structure is induced by a pair of commutative (cyclic) $A_{\infty} / L_{\infty }$ products. }

\vspace{2mm} 

Formulation of superstring field theory has developed with understandings about how we can obtain gauge-invariant operator insertions into string interactions. 
Particularly, in \cite{Goto:2015pqv, Erler:2014eba, Berkovits:1995ab, Erler:2013xta, Berkovits:1998bt, Okawa:2004ii, Berkovits:2004xh, Matsunaga:2013mba, Matsunaga:2014wpa}, gauge invariant actions are constructed by operator insertions using first two of $( \xi (z) , \eta (z) ; \phi (z) )$, fermionic superconformal ghosts. 
Insertions of $\eta (z)$ are very simple because it has conformal weight $1$ and is just a (nilpotent) current: 
On the basis of it, a gauge-invariant action which has a WZW-like form was proposed by Berkovits in an elegant way \cite{Berkovits:1995ab}.  
However, at the same time, these $\eta $-insertions enlarges the gauge symmetry of the theory, and two nilpotent gauge generators appear. 
Insertions of $\xi (z)$ are rather complicated but also possible: 
Using it with nonassociative regulators for \cite{Witten:1986qs, Wendt:1987zh}, Erler, Konopka, and Sachs constructed an $A_{\infty }$ action \cite{Erler:2013xta}. 
This theory does not necessitate to extend gauge symmetry. 
However, to be gauge invariant, a state $\Phi $ appearing in the action must satisfy the constraint equation: $\oint  \eta (z) \, \Phi = 0$\,. 

\vspace{2mm} 

In the NS-NS sector, the situation becomes somewhat complicated: 
There exist three nilpotent generators of gauge transformations, and we have to impose two constraint equations. 
To see this extended gauge symmetry, let us recall the kinetic term of an NS-NS action, which was given by Berkovits based on his $\mathcal{N}=4$ topological prescription\cite{Berkovits:1998bt}, 
\begin{subequations}
\begin{align} 
\label{free large}
S [\Psi ] = - \frac{1}{2} \big{\langle } \, \Psi \, , \, Q \, \eta \, \teta \, \Psi \, \big{\rangle }  + \dots \, , 
\end{align}
where $Q$ is the BRST operator and $\langle A, B \rangle \equiv \langle A | c_{0}^{-} | B \rangle $ is the BPZ inner product with $c_{0}^{-} \equiv \frac{1}{2} (c_{0} - \tilde{c}_{0})$-insertion. 
An NS-NS string field $\Psi$ is total ghost number $0$, left-moving picture number $0$, right-moving picture number $0$ state in the left-and-right large Hilbert space.\footnote{In this paper, we often call the state space whose superconformal ghost sector is spanned by $( \xi (z) , \eta (z) ; \phi (z) )$ and $(\tilde{\xi } (\tilde{z}) , \teta (\tilde{z}) ; \tilde{\phi }(\tilde{z}))$ as the left-and-right large Hilbert space $\cH$. Likewise, we call the state space consists of states belonging to the kernels of both $\eta $ and $\teta $ as the small Hilbert space $\cH _{\rm S}$. We always impose $(b_{0}-\tilde{b}_{0}) \Psi = (L_{0} - \tilde{L}_{0} ) \Psi = 0$ for all closed superstring field $\Psi$.} 
We write $\eta$, $\teta$, $\xi $, and $\tilde{\xi }$ for the zero modes of $\eta (z)$, $\teta (\tilde{z})$,  $\xi (z)$, and $\tilde{\xi } (\tilde{z})$, respectively. 
As one expects from its construction, it is invariant under the gauge transformations 
\begin{align}
\delta \Psi = \eta \, \Omega + \teta \, \widetilde{\Omega } + Q \, \Lambda + \dots \, , 
\end{align}
\end{subequations} 
where $\Omega , \widetilde{\Omega }$, and $\Lambda $ denote gauge parameter fields. 
We thus have three nilpotent gauge generators. 
When we include all interacting terms, three nonlinear extensions of these nilpotent generators appear\cite{Matsunaga:2013mba, Matsunaga:2014wpa, Goto:2015pqv}. 
Then, a full action has a Wess-Zumino-Witten-like form. 
To see constraints, it is helpful to consider the kinetic term\footnote{Note that these two free actions are equivalent each other with linear partial gauge fixing or trivial up-lift. For example, recall that $\Psi $ of (\ref{free large}) is obtained by an embedding of $\Phi $ of (\ref{free small}) such as $\eta \, \teta \, \Psi = \Phi $. } of the $L_{\infty }$ action\cite{Erler:2014eba}, 
\begin{subequations} 
\begin{align} 
\label{free small}
S [\Phi ] = - \frac{1}{2} \big{\langle } \, \xi \, \tilde{\xi } \, \Phi \, , \, Q \, \Phi \, \big{\rangle }  + \dots \, . 
\end{align}
An NS-NS string field $\Phi $ is total ghost number $2$, left-moving picture number $-1$, and right-moving picture number $-1$ state satisfying two constraint equations: $\eta \, \Phi = 0$ and $\teta \, \Phi = 0$. 
One can find that if and only if $\Phi $ satisfies constraints, the action has gauge invariance under 
\begin{align}
\delta \Phi = Q \, \lambda + \dots \, , 
\end{align}
\end{subequations} 
where the gauge parameter $\lambda $ also satisfies constraints: $\eta \, \lambda = 0$ and $\teta \, \lambda = 0$\,. 
In \cite{Erler:2014eba}, starting from Zwiebach's bosonic string products \cite{Zwiebach:1992ie} and finding appropriate gauge invariant $(\xi ; \tilde{\xi })$-insertions, they constructed a suitable NS-NS string products which satisfy (cyclic) $L_{\infty }$ relations, 
\begin{align*}
{\bf L}^{\rm NS,NS} : \hspace{3mm} Q \, , \hspace{3mm} 
L_{2} ( \hspace{1mm} \boldsymbol{\cdot } \hspace{1mm} , \hspace{1mm} \boldsymbol{\cdot } \hspace{1mm} ) \, , \hspace{3mm} 
L_{3} ( \hspace{1mm} \boldsymbol{\cdot } \hspace{1mm} , \hspace{1mm} \boldsymbol{\cdot } \hspace{1mm} , \hspace{1mm} \boldsymbol{\cdot } \hspace{1mm} ) \, , \hspace{3mm} 
L_{4} ( \hspace{1mm} \boldsymbol{\cdot } \hspace{1mm} , \hspace{1mm} \boldsymbol{\cdot } \hspace{1mm} , \hspace{1mm} \boldsymbol{\cdot } \hspace{1mm} , \hspace{1mm} \boldsymbol{\cdot } \hspace{1mm} ) \, , \,\,\, \dots \, ,
\end{align*}
and gave a full action whose interacting terms satisfy $L_{\infty }$ relations. 
When we include all interactions, to be gauge invariant (or to be cyclic), a state $\Phi $ appearing in the $L_{\infty }$ action must satisfy two constraint equations: $\eta \, \Phi = 0$ and $\teta \, \Phi = 0$\,. 
From these analysis, we achieve an idea that a triplet of three nilpotent objects determines the gauge structure of the NS-NS theory: 
By identifying two of them as constraints, one can construct a gauge invariant action. 

\vspace{2mm}

Actually, on the basis of this idea, one can generalise or rephrase the construction of the $L_{\infty }$ action as follows. 
Let $\varphi $ be a dynamical string field. 
We first consider a state $\Phi _{\eta \teta } [ \varphi ]$, which will be a functional of $\varphi $, satisfying two constraint equations, 
\begin{align*}
\boldsymbol{\eta } \, \Phi _{\eta \teta } [ \varphi ] & = 0 \, ,
\\
\boldsymbol{\teta } \, \Phi _{\eta \teta } [ \varphi ] & = 0 \, .
\end{align*}
Then, using this $\Phi _{\eta \teta } [ \varphi ]$, a gauge invariant action whose on-shell condition is given by 
\begin{align*}
Q \, \Phi _{\eta \teta } [\varphi ] + \sum_{n=2}^{\infty } \frac{1}{n!} L_{n} \big( \overbrace{\Phi _{\eta \teta } [\varphi ] \, , \, \dots \, , \, \Phi _{\eta \teta } [\varphi ] }^{n} \big) = 0 
\end{align*}
can be constructed: 
All properties we need are derived from constraint equations for $\Phi _{\eta \teta } [\varphi ]$. 
The resultant action has a WZW-like form and one can prove its gauge invariance via a WZW-like manner without using specific properties of $\varphi $. 
As we will see in section 5, by taking $\varphi = \Phi$ of (\ref{free small}), it reduces to the original $L_{\infty }$ action of \cite{Erler:2014eba}. 
Namely, $L_{\infty }$ formulation is completely described by a triplet of $L_{\infty }$ product $(\boldsymbol{\eta } , \boldsymbol{\teta } \, ; {\bf L}^{\rm NS,NS} )$. %and has the WZW-like structure. 
Likewise, every known actions for NS-NS superstring field theory potentially have their WZW-like forms described by their $L_{\infty }$ triplets. 
For the most general form of the WZW-like structure and action, see section 6 and appendix A. 

\vspace{2mm} 

Furthermore, there exist a dual triplet for this $(\boldsymbol{\eta } , \boldsymbol{\teta } \, ; {\bf L}^{\rm NS,NS} )$, which has the completely same information about the gauge structure of the NS-NS theory. 
Using this dual triplet, one can construct alternative WZW-like action, which is our main focus. 
First, in section 2, we find that the NS-NS superstring product ${\bf L}^{\rm NS,NS}$ has two dual $L_{\infty }$ products: 
\begin{align*}
{\bf L}^{\boldsymbol{\alpha }} &: \hspace{3mm} \alpha \, , \hspace{3mm} 
[ \hspace{1.5mm} \boldsymbol{\cdot } \hspace{1.5mm} , \hspace{1.5mm} \boldsymbol{\cdot } \hspace{1.5mm} ]^{\alpha } \, , \hspace{3mm}  
[ \hspace{1.5mm} \boldsymbol{\cdot } \hspace{1.5mm} , \hspace{1.5mm} \boldsymbol{\cdot } \hspace{1.5mm} , \hspace{1.5mm} \boldsymbol{\cdot } \hspace{1.5mm} ] ^{\alpha } \, , \hspace{3mm} 
[ \hspace{1.5mm} \boldsymbol{\cdot } \hspace{1.5mm}, \hspace{1.5mm} \boldsymbol{\cdot } \hspace{1.5mm} , \hspace{1.5mm} \boldsymbol{\cdot } \hspace{1.5mm} , \hspace{1.5mm} \boldsymbol{\cdot } \hspace{1.5mm} ]^{\alpha } \, , \,\,\, \dots  \hspace{3mm} 
(\, \alpha = \, \eta \, , \, \teta \, ) \, . 
%{\bf L}^{\boldsymbol{\eta }} &: \hspace{3mm} \eta \, , \hspace{3mm} 
%[ \hspace{1.5mm} \boldsymbol{\cdot } \hspace{1.5mm} , \hspace{1.5mm} \boldsymbol{\cdot } \hspace{1.5mm} ]^{\eta } \, , \hspace{3mm}  
%[ \hspace{1.5mm} \boldsymbol{\cdot } \hspace{1.5mm} , \hspace{1.5mm} \boldsymbol{\cdot } \hspace{1.5mm} , \hspace{1.5mm} \boldsymbol{\cdot } \hspace{1.5mm} ] ^{\eta } \, , \hspace{3mm} 
%[ \hspace{1.5mm} \boldsymbol{\cdot } \hspace{1.5mm}, \hspace{1.5mm} \boldsymbol{\cdot } \hspace{1.5mm} , \hspace{1.5mm} \boldsymbol{\cdot } \hspace{1.5mm} , \hspace{1.5mm} \boldsymbol{\cdot } \hspace{1.5mm} ]^{\eta } \, , \,\,\, \dots 
\end{align*} 
We will see that as well as $\boldsymbol{\eta }$, $\boldsymbol{\teta }$, or ${\bf L}^{\rm NS,NS}$, these $L_{\infty }$ products have nice algebraic properties. 
Then, one can consider the constraint equations provided by these ${\bf L}^{\boldsymbol{\eta }}$ and ${\bf L}^{\boldsymbol{\teta }}$\,: 
\begin{align*}
&\eta \, \Psi _{\eta \teta } [\varphi ] + \sum_{n=1}^{\infty } \frac{1 }{n!} \big[ \overbrace{ \Psi _{\eta \teta } [\varphi ] \, , \, \dots \, , \, \Psi _{\eta \teta } [\varphi ] }^{n} \big] ^{\eta }  = 0 \, ,
\\ 
&\teta \, \Psi _{\eta \teta } [\varphi ] + \sum_{n=1}^{\infty } \frac{1 }{n!} \big[ \overbrace{ \Psi _{\eta \teta } [\varphi ] \, , \, \dots \, , \, \Psi _{\eta \teta } [\varphi ] }^{n} \big] ^{\teta }  = 0  \, .
\end{align*} 
Using a state $\Psi _{\eta \teta } [\varphi ]$ satisfying these constraint equations, which will be a functional of some dynamical string field $\varphi $, we construct a gauge invariant action whose on-shell condition is 
\begin{align*}
{\bf Q} \, \Psi _{\eta \teta } [\varphi ] = 0 \, . 
\end{align*}
It also has a WZW-like form and one can prove its gauge invariance without details of $\varphi $, which we explain in section 3. 
The $L_{\infty }$ triplet $({\bf L}^{\boldsymbol{\eta }} , {\bf L}^{\boldsymbol{\teta }} \, ; {\bf Q} )$ determines this WZW-like structure and action. 
All necessitated properties can be derived from the constraint equations for $\Psi _{\eta \teta } [\varphi ]$, and we give two explicit forms of this key functional $\Psi _{\eta \teta } [\varphi ]$ in section 4. 
As we show in section 5, these WZW-like actions described by $({\bf L}^{\boldsymbol{\eta }} , {\bf L}^{\boldsymbol{\teta }} \, ; {\bf Q})$ and $(\boldsymbol{\eta } , \boldsymbol{\teta } \,; {\bf L}^{\rm NS,NS})$ are off-shell equivalent, which would be an interesting aspect of the WZW-like structure. 
Through these analysis, we would like to show that a triplet of mutually commutative $L_{\infty }$ products completely determines the WZW-like structure of NS-NS superstring field theory, which is our main result. 

\vspace{2mm} 

In section 5, we present detailed properties of our WZW-like action. 
Firstly, we show that as well as that of the NS sector, our WZW-like action of the NS-NS sector has a single functional form which consists of single functionals $\Psi _{\eta \teta } [\varphi ]$ and elementally operators. 
Secondly, using this single functional form, we prove the equivalence of two constructions given in section 4. 
Thirdly, we clarify the relation to $L_{\infty }$ theory: 
We find that our WZW-like action and the $L_{\infty }$ action are off-shell equivalent. 
Then we give a short discussion about off-shell duality of equivalent $L_{\infty }$ triplets. 
Finally, we discuss the relation to the earlier WZW-like theory proposed by \cite{Matsunaga:2014wpa}. 
With a brief summary of the WZW-like structure, we end with conclusion in section 6. 
In appendix A, we discuss the WZW-like action based on a general (nonlinear) $L_{\infty }$ triplet $(\bL ^{c} , \bL ^{\tilde{c}} \,; \bL ^{p})$\,. 
We show that as well as other known WZW-like actions, it also satisfies the expected properties. 

%\clearpage

\section{Two triplets of $L_{\infty }$} 

In this section, we present two triplets of mutually commutative $L_{\infty }$ products. 
The $L_{\infty }$ triplet completely determines the WZW-like action: its form, gauge structure and all algebraic properties. 
As we will see, it gives the most fundamental ingredient of NS-NS superstring field theory because every known actions potentially have the WZW-like form. 

We write the graded commutator of two co-derivations $\boldsymbol{D}_{1}$ and $\boldsymbol{D}_{2}$ as 
\begin{align*}
\Ld \, \boldsymbol{D}_{1} , \, \boldsymbol{D}_{2} \, \Rd \equiv \, \boldsymbol{D}_{1} \,\, \boldsymbol{D}_{2} -(-)^{\boldsymbol{D}_{1} \boldsymbol{D}_{2} } \boldsymbol{D}_{2} \,\, \boldsymbol{D}_{1} \, . 
\end{align*}
Note that it satisfies Jacobi identity exactly (without $L_{\infty }$ homotopy terms): 
\begin{align*}
\Ld \, \boldsymbol{D}_{1} , \, \ld \, \boldsymbol{D}_{2} \, , \, \boldsymbol{D}_{3} \, \rd \Rd  
+ (-)^{\boldsymbol{D}_{1} ( \boldsymbol{D}_{2} + \boldsymbol{D}_{3} ) } \Ld \, \boldsymbol{D}_{1} , \, \ld \, \boldsymbol{D}_{2} , \, \boldsymbol{D}_{3} \, \rd \Rd  
+ (-)^{\boldsymbol{D}_{3} ( \boldsymbol{D}_{1} + \boldsymbol{D}_{2} ) } \Ld \, \boldsymbol{D}_{1} , \, \ld \, \boldsymbol{D}_{2} , \, \boldsymbol{D}_{3} \, \rd \Rd 
= 0 \, . 
\end{align*}

\niu{Original $L_{\infty }$ triplet\,: $(\boldsymbol{\eta } , \boldsymbol{\teta } \, ; {\bf L}^{\rm NS,NS} )$}

As we explained, the constraint equations and the of-shell condition of the $L_{\infty }$ action is described by a triplet of mutually commutative $L_{\infty }$-products $(\boldsymbol{\eta } , \boldsymbol{\teta } \, ; {\bf L}^{\rm NS,NS} )$, which is the first one of two $L_{\infty }$ triplets. 
The other $L_{\infty }$ triplet is its dual and has the completely same information. 
Before considering its dual, let us recall how this $L_{\infty }$ product ${\bf L}^{\rm NS,NS}$ was constructed. 
In \cite{Erler:2014eba}, they introduced a generating function ${\bf L} ( s,\ts \, ; t)$ for a series of $L_{\infty }$ products, and required that ${\bf L} ( 0 , 0 ; 0 ) \equiv {\bf Q}$ and ${\bf L} ( 1,1 \,; 0 )$ gives Zwiebach's string products of bosonic closed string field theory \cite{Zwiebach:1992ie}. 
To relate this ${\bf L} ( s,\ts \,; t)$ with operator insertions, it is helpful to consider another generating function $\boldsymbol{\mu } ( s,\ts \,; t)$ which has all information about operator insertions and will implicitly determine the gauge invariance.  
They called this $\boldsymbol{\mu } ( s , \ts \, ; t)$ as  a gauge product. 
The NS-NS $L_{\infty }$ products ${\bf L}^{\rm NS,NS}$ is included in this generating function ${\bf L}(s,\ts \, ; t)$. 
By imposing or solving the recursive equations, 
\begin{align*}
\frac{\partial }{\partial s} \, {\bf L} ( s , \ts \, ; t ) & = \Ld \, \boldsymbol{\eta } \, , \, \boldsymbol{\mu } ( s , \ts \, ; t ) \, \Rd \, , 
\hspace{5mm} 
\frac{\partial }{\partial \ts } \, {\bf L} ( s , \ts \, ; t )  = \Ld \, \boldsymbol{\teta } \, , \, \boldsymbol{\mu } ( s , \ts \, ; t ) \, \Rd \, ,
\end{align*} 
with the initial conditions, one can obtain an appropriate ${\bf L}(s,\ts \,; t)$ from $\boldsymbol{\mu }(s,\ts \,; t)$, and vice versa. 
This ${\bf L}(s,\ts \,; t)$ is a series of $L_{\infty }$ products with operator insertions satisfying $\ld \, {\bf L}\, , \boldsymbol{\eta } \, \rd = 0$ and $\ld \, {\bf L}\, , \boldsymbol{\teta } \, \rd = 0$\,. 
As shown in \cite{Erler:2014eba}, explicit forms of ${\bf L}(s,\ts \,; t)$ and $\boldsymbol{\mu } (s,\ts \,; t)$ can be determined by solving the recursive equation, 
\begin{align*} 
\frac{\partial }{\partial t} \, {\bf L} ( s , \ts \, ; t ) & = \Ld \, {\bf L} ( s, \ts \, ; t ) \, , \, \boldsymbol{\mu } ( s , \ts \, ; t ) \, \Rd  \, , 
\end{align*}
which ensures $L_{\infty }$ relations $\ld \, {\bf L} \, , {\bf L} \, \rd = 0$\,. 
Using these ${\bf L} ( s , \ts \, ; t )$ and $\boldsymbol{\mu } ( s , \ts \, ; t )$\,, the NS-NS superstring $L_{\infty }$ products ${\bf L}^{\rm NS,NS}$ is given by the $s=0$, $\ts = 0$, and $t=1$ value of ${\bf L} (s,\ts \, ; t)$\,: 
\begin{align*}
{\bf L}^{\rm NS,NS} \equiv {\bf L} ( s = 0, \ts = 0 ; t = 1) \, . 
\end{align*} 
We write $L_{n}$ for the $n$-th product of ${\bf L}^{\rm NS,NS} $ as follows, 
\begin{align*}
L_{n} \big( A_{1} \, , \, \dots , \, , A_{n} \big) \equiv \pi _{1} \, 
{\bf L}^{\rm NS,NS} \big( A_{1} \wedge \, \dots \, \wedge A_{n}  \big) \, . 
\end{align*}

\vspace{2mm} 

Note that this $\boldsymbol{\mu } ( s , \ts \, ; t )$ has all information about gauge-invariant operator insertions and thus about how to construct the NS-NS products. 
Once we determine $\boldsymbol{\mu } ( s , \ts \, ; t )$, how to gauge-invariantly insert $\boldsymbol{\xi }$, $\boldsymbol{\tilde{\xi }}$, and picture-changing operators, the NS-NS $L_{\infty }$ product ${\bf L}^{\rm NS,NS}$ is given by the $t = 1$ value solution of the linear differential equation 
\begin{align*}
\frac{\partial }{\partial t} \, {\bf L}^{\rm NS, NS} (t) = \Ld \, {\bf L}^{\rm NS, NS} (t) \, , \, \boldsymbol{\mu }(t) \, \Rd \,  
\end{align*} 
with the initial condition ${\bf L}^{\rm NS,NS} (t = 0) = {\bf Q}$, where  $\boldsymbol{\mu }(t) \equiv \boldsymbol{\mu } ( s= 0 , \ts = 0 ; t )$. 
Hence, we can solve it by iterated integration (with direction) and have the following expression, 
\begin{align*}
{\bf L}^{\rm NS,NS} = \overset{\rightarrow}{\cP } \exp \Big[ - \int_{0}^{t} dt \, \boldsymbol{\mu } (t) \Big] \, {\bf Q} \, \, \overset{\leftarrow}{\cP } \exp \Big[ \int_{0}^{t} dt \, \boldsymbol{\mu } (t)  \Big]   \, . 
\end{align*}
For brevity, we write $\widehat{\bf G}$ for this iterated integral with direction and write ${\bf L}^{\rm NS,NS} = \widehat{\bf G}^{-1} \, {\bf Q} \, \widehat{\bf G}$\,:
\begin{align*}
\widehat{\bf G} \equiv \overset{\leftarrow}{\cP } \exp \Big[ \int_{0}^{t} dt \, \boldsymbol{\mu } (t)  \Big] \, . 
\end{align*}
It is a path-ordered exponential of coderivation $\boldsymbol{\mu }$, and thus a natural cohomomorphism of $L_{\infty }$ algebras. 
In this form, $L_{\infty } $ relations look trivial: $({\bf L})^{2} = \widehat{\bf G}^{-1} ( {\bf Q} )^{2}\, \widehat{\bf G} = 0$. 
Using this form, we find two dual $L_{\infty }$ products for ${\bf L}^{\rm NS,NS}$\, and a dual of the $L_{\infty }$ triplet $(\boldsymbol{\eta } , \boldsymbol{\teta } \,; {\bf L}^{\rm NS,NS})$\,. 

\niu{Dual $L_{\infty }$ triplet\,: $({\bf L}^{\boldsymbol{\eta }} , {\bf L}^{\boldsymbol{\teta }} \, ; {\bf Q} )$} 

By construction, the NS-NS product ${\bf L}^{\rm NS,NS}$ commutes with two $L_{\infty }$ products $\boldsymbol{\eta }$ and $\boldsymbol{\teta }$\,: $\ld \, {\boldsymbol \eta } \,, {\bf L}^{\rm NS,NS} \, \rd = 0$ and $\ld \, \boldsymbol{ \teta } \,, {\bf L}^{\rm NS,NS} \, \rd = 0$\,. 
Thus, there exist two dual $L_{\infty }$ products for ${\bf L}^{\rm NS,NS}$. 
Using path-ordered exponential map $\widehat{\bf G}$, one can obtain these dual $L_{\infty }$ products as follows, 
\begin{subequations} 
\begin{align}
\label{dual left}
{\bf L}^{\boldsymbol \eta } & \equiv \widehat{\bf G} \, {\boldsymbol \eta } \, \widehat{\bf G}^{-1} \, , 
\\ \label{dual right} 
{\bf L}^{\boldsymbol{ \teta }} & \equiv \widehat{\bf G} \, \boldsymbol{ \teta } \, \widehat{\bf G}^{-1} \, . 
\end{align}
\end{subequations} 
One can quickly find that these products satisfy $L_{\infty }$ relations $({\bf L}^{\boldsymbol{\alpha }})^{2} = \widehat{\bf G} \, ( \boldsymbol{\alpha } )^{2}\, \widehat{\bf G}^{-1} = 0$ because of $( \boldsymbol{\alpha } )^{2} = 0$ for ${\boldsymbol \alpha } = {\boldsymbol \eta} ,\, \widetilde{\boldsymbol \eta }$\,, and have $Q$-derivation properties
\begin{align*}
{\bf Q} \, {\bf L}^{\boldsymbol \alpha } = \widehat{\bf G} \, ( \widehat{\bf G}^{-1} \, {\bf Q} \, \widehat{\bf G} )\, {\boldsymbol \alpha } \, \widehat{\bf G}^{-1} = - \widehat{\bf G} \, {\boldsymbol \alpha } \, ( \widehat{\bf G}^{-1} \, {\bf Q} \, \widehat{\bf G} ) \, \widehat{\bf G}^{-1} = - {\bf L}^{\boldsymbol \alpha } \, {\bf Q} \,  
\end{align*} 
because of $\ld \, {\bf L}^{\rm NS,NS}, \boldsymbol{\alpha } \, \rd = 0$ for ${\boldsymbol \alpha } = {\boldsymbol \eta} ,\, \widetilde{\boldsymbol \eta }$, which will provide nonlinear extensions of constraint equations. 
Hence, as well as $(\boldsymbol{\eta } , \boldsymbol{\teta } \,; {\bf L}^{\rm NS,NS} )$, the triplet of $L_{\infty }$-products $({\bf L}^{\boldsymbol{\eta }} , {\bf L}^{\boldsymbol{\teta }} \, ; {\bf Q} )$ is nilpotent and mutually commutative. 
Note that we found the correspondence of the commutativity: 
\begin{align}
\label{Duality} 
\Ld \, {\boldsymbol \alpha } \, , \, {\bf L}^{\rm NS,NS} \, \Rd = 0 
\hspace{5mm} \iff \hspace{5mm} 
\Ld \, {\bf L}^{\boldsymbol{\alpha } } \, , \, {\bf Q} \, \Rd = 0 \, , \hspace{5mm} 
(\boldsymbol{\alpha } = \boldsymbol{\eta } , \boldsymbol{\teta } ) \, . 
\end{align} 
It is owing to an invertible cohomomorphism $\widehat{\bf G}$, and thus the $L_{\infty }$ triplet $({\bf L}^{\boldsymbol{\eta }} , {\bf L}^{\boldsymbol{\teta }} \, ; {\bf Q} )$ has the completely same information as $(\boldsymbol{\eta } ,\boldsymbol{\teta } \, ; {\bf L}^{\rm NS,NS} )$. 
We thus call $({\bf L}^{\boldsymbol{\eta }} , {\bf L}^{\boldsymbol{\teta }} \, ; {\bf Q} )$ as the dual $L_{\infty }$ triplet. 
When $\widehat{\bf G}$ is cyclic in the BPZ inner product, this correspondence provides the equivalence of WZW-like actions governed by equivalent $L_{\infty }$ triplets (See section 5.). 
In this paper, we write the $n$-th product of ${\bf L}^{\boldsymbol{\alpha }}$ as follows, 
\begin{align*}
[ A_{1} , \dots , A_{n} ]^{\alpha } := {\pi_1 } \widehat{\bf G} \, {\boldsymbol \alpha } \, \widehat{\bf G} ^{-1} ( A_{1}\wedge \dots \wedge A_{n} ) , \hspace{5mm} ( \alpha = \eta , \, \widetilde{\eta } ). 
\end{align*}

\subsection*{Nilpotent relations and Derivation properties} 

For later use, we present explicit forms of algebraic relations satisfied by $({\bf L}^{\boldsymbol{\eta }} , {\bf L}^{\boldsymbol{\teta }} \, ; {\bf Q} )$ and some details of related properties. 
The dual $L_{\infty }$ product ${\bf L}^{\boldsymbol{\alpha }}$ for $\alpha = \eta , \widetilde{\eta }$ satisfies $L_{\infty }$-relations, $({\bf L}^{\boldsymbol{\alpha }} )^{2} = 0$. 
In terms of the $n$-th component, we have 
\begin{subequations}
\begin{align}
\label{single nilpotent}
\sum_{\sigma }\sum_{k=1}^{n} (-)^{|\sigma |} \big[ [ A_{i_{\sigma (1)}} , \dots , A_{i_{\sigma (k)}} ]^{\alpha } , A_{i_{\sigma (k+1)}} , \dots , A_{i_{\sigma (n) }} \big] ^{\alpha } = 0 \, ,  
\end{align}
where $\sigma $ runs over all possible permutations and $(-)^{|\sigma |}$ denotes the sign of the corresponding permutation. 
Likewise, ${\bf L}^{\boldsymbol{\alpha }} \, {\bf Q} + {\bf Q} \, {\bf L}^{\boldsymbol{\alpha }} = 0$ implies that we have $Q$-derivation properties , 
\begin{align} 
\label{derivation}
Q \big[ A_{1} , \dots , A_{n} \big] ^{\alpha } + \sum_{i=1}^{n-1} (-)^{A_{1} + \dots + A_{k-1}} \big[ A_{1} , \dots, Q A_{k} , \dots , A_{n} \big] ^{\alpha } = 0 \, , 
\end{align}
where the upper index of $(-)^{A}$ means the grading of $A$, namely, the total ghost number of $A$. 
The commutativity ${\bf L}^{\boldsymbol{\eta }} \, {\bf L}^{\boldsymbol{\teta }} + {\bf L}^{\boldsymbol{\teta }} \, {\bf L}^{\boldsymbol{\eta }} = 0$ provides 
\begin{align} 
\label{double nilpotent}
\sum_{\alpha_{1} , \alpha _{2} = \eta , \widetilde{\eta} }\sum_{\sigma }\sum_{k=1}^{n} (-)^{|\sigma |} \big[ [ A_{{\sigma (1)}} , \dots , A_{{\sigma (k)}} ]^{\alpha _{1}} , A_{{\sigma (k+1)}} , \dots , A_{{\sigma (n) }} \big] ^{\alpha _{2}} = 0 \, . 
\end{align}
\end{subequations} 
The lowest relation of (\ref{double nilpotent}) is just $\eta \, \teta + \teta \, \eta = 0$, which would be very familiar. 
One can quickly find that the second lowest relation of (\ref{double nilpotent}) is given by 
\begin{align*}
\eta \, \big[ \, A \, , B \, \big] ^{\teta } + \big[ \, \eta \, A \, , B \, \big] ^{\teta } + (-)^{A} \big[ \, A \, , \eta \, B \, \big] ^{\teta } + \teta \, \big[ \, A \, , B \, \big] ^{\eta } + \big[ \, \teta \, A \, , B \, \big] ^{\eta } +(-)^{A} \big[ \, A \, , \teta \, B \, \big] ^{\eta } = 0 \, , 
\end{align*}
which is the matching of (crossed) Leibniz rules. 
Similarly, one can derive any higher relations of (\ref{double nilpotent}). 
It may look a little complicated, but it is powerful and exact. 

\vspace{2mm} 

\niu{Maurer-Cartan element and Shifted $L_{\infty }$} 

There is a special element of the $L_{\infty }$ algebra of ${\bf L}^{\boldsymbol{\alpha }}$ for $\alpha = \eta , \teta $, 
\begin{align*}
\mathcal{MC}_{L^{\alpha }} ( A ) \equiv \alpha \, A + \sum_{n=1}^{\infty } \frac{1}{n!} \big[ \, \overbrace{A \, , \, \dots , A}^{n}\, \big] ^{\alpha } , 
\end{align*}
which we call the Maurer-Cartan element for ${\bf L}^{\boldsymbol{\alpha }}$. 
As we will see, this element plays central role in WZW-like theory: 
It appears in the constraint equations, in the on-shell condition, and in the WZW-like action. 
Likewise, we often refer $\mathcal{MC}_{Q} (A) \equiv Q A $ as the Maurer-Cartan element for ${\bf Q}$\,. 
There is an natural operation, a shift of the products, in $L_{\infty }$ algebras. 
For any state $A$, the $A$-shifted products are defined by 
\begin{align*}
\big[ \, B_{1} \, , \dots , B_{n} \, \big] ^{\alpha }_{A} \equiv \sum_{n=0}^{\infty } \frac{1}{n!} \big[ \overbrace{A \, , \dots , A}^{n} \, , B_{1} \, , \dots \, , B_{n} \, \big] ^{\alpha } \, . 
\end{align*} 
Note that the Maurer-Cartan element $\mathcal{MC}_{L^{\alpha }} (A)$ behaves as the $A$-shifted $0$-th product. 
One can check that with $\mathcal{MC}_{L^{\alpha }}(A)$, the $A$-shifted products satisfy weak $L_{\infty }$ relations: 
\begin{subequations}
\begin{align}
\sum_{\sigma } \sum_{k=1}^{n} (-)^{|\sigma |} \Big[  \big[ B_{\sigma (1)} , \dots , B_{\sigma (k)} \big] ^{\alpha }_{A} , B_{\sigma (k+1)} , \dots , B_{\sigma (n)} \Big] ^{\alpha }_{A} = - \big[ \mathcal{MC}_{L^{\alpha }} (A) , B_{1} , \dots , B_{n} \big] ^{\alpha }_{A} . \label{weak L1}
\end{align}
It implies that when given state $A$ satisfies the Maurer-Cartan equation $\mathcal{MC}_{L^{\alpha }}(A) = 0$, then the $A$-shifted products exactly satisfy the $L_{\infty }$ relations. 
Similarly, one can consider the shift of (\ref{double nilpotent}) and obtain the weakly commuting relations of two $A$-shifted products: 
\begin{align} 
\sum_{\alpha_{1} , \alpha _{2} = \eta , \widetilde{\eta} } & \sum_{\sigma }\sum_{k=1}^{n} (-)^{|\sigma |} \Big[ \big[ \, B_{{\sigma (1)}} \, , \dots , B_{{\sigma (k)}} \, \big] ^{\alpha _{1}}_{A} \, , \, B_{{\sigma (k+1)}} \, , \dots , B_{{\sigma (n) }} \Big] ^{\alpha _{2}}_{A}  
\no & \hspace{10mm} 
= - \big[ \, \mathcal{MC}_{L^{\eta }} (A) \, , B_{1} \, , \dots , B_{n} \, \big] ^{\teta }_{A} 
- \big[ \, \mathcal{MC}_{L^{\teta }} (A) \, , B_{1} \, , \dots , B_{n} \, \big] ^{\eta }_{A} . \label{weak L2}
\end{align}
\end{subequations} 
We thus find that two $A$-shifted products commute if and only if given state $A$ satisfies both of the Maurer-Cartan equations $\mathcal{MC}_{L^{\eta }} (A) = 0$ and $\mathcal{MC}_{L^{\teta }} (A) = 0$\,. 
Using these relations, we prove the gauge invariance of the WZW-like action for NS-NS superstring field theory.

%%%%%%%%%%%%%%%%
%\clearpage

\section{WZW-like action}

Once we have a triplet of mutually commutative $L_{\infty }$-products $( {\bf L}^{\boldsymbol{\eta }} , {\bf L}^{\boldsymbol{\teta }} \, ; {\bf Q} )$, by using these to provide constraints or on-shell equations, we can construct a gauge invariant action, which we explain in this section. 
We would like emphasis that one can achieve the gauge invariance without using detailed properties of a dynamical string field of the theory. 
All we need are two functional fields and their algebraic relations. 

\subsection*{Algebraic Ingredients} 

In our WZW-like formulation of the NS-NS sector, two $L_{\infty }$-products ${\bf L}^{\boldsymbol{\eta }}$ and ${\bf L}^{\boldsymbol{\teta }}$ are used to define constraint equations for (functional) fields, the other $L_{\infty }$-product ${\bf Q}$ is used to give the on-shell condition, and their mutual commutativity ensures the gauge invariance. 

A functional field $\Psi _{\eta \eta }[ \varphi ]$ satisfying these constraint equations plays the most important role, which we call {\it a pure-gauge-like (functional) field}. 
With this functional $\Psi _{\eta \teta }[\varphi ]$, the commutativity of $L_{\infty }$-products induces key algebraic relations, {\it WZW-like relations}. 
They make possible to prove the gauge invariance without details of the dynamical string field $\varphi $ of the theory. 

\niu{WZW-like functional field}

Let $\Psi _{\eta \widetilde{\eta }} = \Psi _{\eta \teta } [\varphi ]$ be a Grassmann even, ghost number $2$, left-moving picture number $-1$, and right-moving picture number $-1$ state in the left-and-right large Hilbert space: $\eta \, \teta \, \Psi _{\eta \teta } \not= 0$. 
We call this $\Psi _{\eta \teta }$ {\it a pure-gauge-like (functional) field} when $\Psi _{\eta \teta }$ satisfies the constraint equations: 
\begin{subequations}
\begin{align}
&\eta \, \Psi _{\eta \widetilde{\eta }} + \sum_{n=1}^{\infty } \frac{1 }{n!} \big[ \overbrace{ \Psi _{\eta \widetilde{\eta }} \, , \dots , \Psi _{\eta \widetilde{\eta }} }^{n} \big] ^{\eta }  = 0 , \label{WZW 1a} 
\\ %\hspace{5mm} 
&\teta \, \Psi _{\eta \widetilde{\eta }} + \sum_{n=1}^{\infty } \frac{1 }{n!} \big[ \overbrace{ \Psi _{\eta \widetilde{\eta }} \, , \dots , \Psi _{\eta \widetilde{\eta }}  }^{n} \big] ^{\teta }  = 0 . \label{WZW 1b}
\end{align} 
\end{subequations}
In other words, $\Psi _{\eta \teta } [\varphi ]$ gives a solution of the Maurer-Cartan equations for the both dual products (\ref{dual left}) and (\ref{dual right}). 
Therefore, two $\Psi _{\eta \teta }[\varphi ]$-shifted products again have $L_{\infty }$ relations and commute each other. 
One can define two linear operators $D_{\eta }$ and $D_{\teta }$ acting on any state $A$ by 
\begin{align*}
D_{\alpha } A \equiv \alpha \, A + \sum_{n=1}^{\infty } \frac{1}{n!} \big[ \, \overbrace{\Psi _{\eta \widetilde{\eta }} \, , \dots , \Psi _{\eta \widetilde{\eta }} }^{n} \, , A \, \big] ^{\alpha } , \hspace{5mm}  ( \alpha = \eta , \widetilde{\eta }  ) ,
\end{align*}
and two bilinear products of any states $A$ and $B$ by 
\begin{align*} 
\big[ \, A \, ,  B \, \big] ^{\alpha }_{\Psi _{\eta \teta }} \equiv \big[ \, A \, , B \, \big] ^{\alpha } + \sum_{n=1}^{\infty } \frac{1}{n!} \big[ \, \overbrace{\Psi _{\eta \teta } \, , \dots , \Psi _{\eta \teta } }^{n} \, , A \, , B \, \big] ^{\alpha } , \hspace{5mm}  ( \alpha = \eta , \widetilde{\eta }  ) .
\end{align*}
Then, as the first identity of (\ref{weak L1}), one can quickly find that $D_{\eta }$ and $D_{\teta }$ are nilpotent, 
\begin{subequations}
\begin{align}
( D_{\alpha } )^{2} A & = 0 \, ,\hspace{5mm}  ( \alpha = \eta , \widetilde{\eta }  ) . \label{shifted nilpotent}
\end{align}
As the second identity of (\ref{weak L1}), the bilinear product satisfies Liebniz rules, 
\begin{align} 
D_{\alpha } \big[ A \, , B \big] ^{\alpha }_{\Psi _{\eta \teta }} + \big[ D_{\alpha } A \, , B \big] ^{\alpha }_{\Psi _{\eta \teta }} + (-)^{A} \big[ A \, , D_{\alpha } B \big] ^{\alpha }_{\Psi _{\eta \teta }} & = 0 . \label{Leibniz}
\end{align}
Likewise, as the first identity of (\ref{weak L2}), we have the (anti-) commutation relation, 
\begin{align}
\big( \, D_{\eta } \, D_{\teta } + D_{\teta } \, D_{\eta } \, \big) \, A = 0 , \label{anti}
\end{align} 
and as the second identity of (\ref{weak L2}), we can find matching of crossed Liebniz rules, 
\begin{align}
D_{\eta } \, \big[ \, A \, , \, B \, \big] ^{\teta }_{\Psi _{\eta \teta }} + \big[ \, D_{\eta } \, A \, , \, B \, \big] ^{\teta }_{\Psi _{\eta \teta }} + (-)^{A} \big[ \, A \, , \, D_{\eta } \, B \, \big] ^{\teta }_{\Psi _{\eta \teta }} 
\hspace{20mm} & \no 
+ D_{\teta } \, \big[ \, A \, , \, B \, \big] ^{\eta }_{\Psi _{\eta \teta }}  +  \big[ \, D_{\teta } \, A \, , \, B \, \big] ^{\eta }_{\Psi _{\eta \teta }} + (-)^{A} \big[ \, A \, , \, D_{\teta } \, B \, \big] ^{\eta }_{\Psi _{\eta \teta }} & = 0 \, . \label{matching}
\end{align}
\end{subequations} 

\niu{WZW-like relations} 

Let $\mathbf{D}$ be a derivation operator for both $L_{\infty }$-products ${\bf L}^{\boldsymbol{\eta }}$ and ${\bf L}^{\boldsymbol{\teta }}$\,: 
Namely, 
\begin{align*}
(-)^{\mathbf{D} } \, \mathbf{D} \, \big[ A_{1} , \dots , A_{n} \big] ^{\alpha } = \sum_{k=1}^{n} (-)^{\mathbf{D} ( A_{1} + \dots + A_{k-1}) } \big[ A_{1} , \dots, \mathbf{D} \, A_{k} , \dots , A_{n} \big] ^{\alpha }\, ,  \hspace{5mm} ( \alpha = \eta , \widetilde{\eta } )  
\end{align*}
holds for any states $A_{1} , \dots , A_{n} \in \cH$\,. 
For example, since the BRST operator $Q$, a partial differential $\partial _{t}$ with respect to any formal parameter $t \in \mathbb{R}$, and the variation $\delta $ of the dynamical string field satisfy the Leibniz rule for these $L_{\infty }$-products ${\bf L}^{\boldsymbol{\eta }}$ and ${\bf L}^{\boldsymbol{\teta }}$, one can take $\mathbf{D} = Q$, $\partial _{t} $, or $\delta $. 
By acting this $\mathbf{D}$ on the constraint equations (\ref{WZW 1a}) and (\ref{WZW 1b}), we find $D_{\eta } ( \mathbf{D} \, \Psi _{\eta \teta } ) = 0$ and $D_{\teta } ( \mathbf{D} \, \Psi _{\eta \teta } ) = 0$. 
Nilpotent properties $( D_{\eta } )^{2} = 0$ and $(D_{\teta } )^{2} = 0$ imply that with some (functional) state $\Psi _{\mathbf{D} } [\varphi ]$ belonging to the left-and-right large Hilbert space $\cH $, we have  
\begin{align}
\label{WZW 2} 
- (-)^{\mathbf{D} } \, \mathbf{D} \, \Psi _{\eta \widetilde{\eta }} [\varphi ] \, = D_{\eta } \, D_{\teta } \, \Psi _{\mathbf{D} } [\varphi ] \, , 
\end{align}
which is the most important relation in the WZW-like formulation of the NS-NS sector, {\it the WZW-like relation}. 
Note that the existence of the (functional) state $\Psi _{\mathbf{D}} [\varphi ]$ is ensured by the fact\footnote{In section 5, we will see this fact again.} that both $D_{\eta }$-complex and $D_{\teta }$-complex are exact in the left-and-right large Hilbert space $\cH$\,. 
We call this $\Psi _{\mathbf{D}} [\varphi ]$ satisfying (\ref{WZW 2}) as {\it an associated (functional) field}. 

When the derivation operator $\mathbf{D}$ has ghost number $g$, left-moving picture number $p$, and right-moving picture $\widetilde{p}$, the associated field $\Psi _{\mathbf{D} } [\varphi ]$ has the same quantum numbers: 
Its ghost number is $g$, left-moving picture number is $p$, and right-moving picture number is $\widetilde{p}$. 

\vspace{2mm} 

We started with the $L_{\infty }$ triplet $( {\bf L}^{\boldsymbol{\eta }} , {\bf L}^{\boldsymbol{\teta }} \, ; {\bf Q} )$ and obtained the above algebraic ingredients by using two of it as constraints of theory. 
What is the use of the last $L_{\infty }$? 
As we will see, its Maurer-Cartan equation gives a constraint describing the mass shell with the above $\Psi _{\eta \teta } [\varphi ]$\,: 
\begin{align}
\label{MC Q} 
Q \, \Psi _{\eta \teta } [\varphi ] = 0 \, . 
\end{align}
Note that this (\ref {MC Q}) is also a special case of (\ref{WZW 2}). 
Thus, the above three relations $(\ref{WZW 1a})$, $(\ref{WZW 1b})$, and $(\ref{WZW 2})$ are fundamental, and we often call them as {\it Wess-Zumino-Witten-like relations} in NS-NS superstring field theory. 

\subsection*{Action, Equations of motion, and Gauge Invariances}

Let $\varphi $ be a dynamical NS-NS string field and $\varphi (t)$ be a path satisfying $\varphi (0) = 0$ and $\varphi (1) = \varphi $, where $t \in [0,1]$ is a real parameter. 
Once we obtain $\Psi _{\eta \widetilde{\eta }} [ \varphi ]$ and $\Psi _{\mathbf{D} } [\varphi ]$ as functionals of given dynamical string field $\varphi $, we can construct a WZW-like action for NS-NS string field theory: 
\begin{align}
\label{Action}
S_{\eta \tilde{\eta }} [\varphi ] = \int _{0}^{1} dt \, \big{\langle } \Psi _{t} [\varphi (t) ] , \, Q \, \Psi _{\eta \tilde{\eta } } [\varphi (t) ] \big{\rangle } , 
\end{align}
where $\Psi _{t} [\varphi (t) ]$ denotes $\Psi _{\mathbf{D} } [\varphi (t)]$ with $\mathbf{D} = \partial _{t}$, the $t$-differential associated (functional) field. 
As we will see, using the variational associated (functional) field $\Psi _{\mathbf{D} }[\varphi ]$ with $\mathbf{D} = \delta $,  the variation of this action is given by $t$-independent form: 
\begin{align}
\delta S_{\eta \tilde{\eta }} [\varphi ] =  \big{\langle } \Psi _{\delta } [\varphi ] , \, Q \, \Psi _{\eta \tilde{\eta } } [\varphi ] \big{\rangle } \label{topo}. 
\end{align}
Then, the WZW-like relation (\ref{WZW 2}) implies that the gauge transformations are given by 
\begin{align}
\Psi _{\delta } [\varphi ] = D_{\eta } \, \Omega + D_{\widetilde{\eta }} \, \widetilde{\Omega } + Q \, \Lambda . 
\end{align}
The equation of motion is given by $t$-independent form 
\begin{align}
 Q \, \Psi _{\eta \widetilde{\eta }} [\varphi ] = D_{\eta } \, D_{\teta } \, \Psi _{Q} [\varphi ] = - D_{\teta } \, D_{\eta } \, \Psi _{Q} [\varphi ] = 0 . 
\end{align}
One can quickly find these facts by using only WZW-like relations, (\ref{WZW 1a}), (\ref{WZW 1b}), and (\ref{WZW 2}), which we explain in the rest.\footnote{These computations are similar to those of the earlier WZW-like action \cite{Matsunaga:2014wpa}.}

\subsection*{Variation of the action} 

Let us recall basic properties of $L_{\infty }$-products and the BPZ inner product. 
The inner product $\langle A , B \rangle$ includes the $c_{0}^{-}$-insertion.\footnote{In the left-and-right large Hilbert space, the inner product $\langle A , B \rangle $ vanishes unless the sum of $A$'s and $B$'s total ghost, left-moving picture, and right-moving picture numbers are $3$, $-1$, and $-1$, respectively.} 
Hence, for $\mathbf{D'} = D_{\eta }$, $D_{\teta }$, or $Q$, we have\footnote{The prime denotes that we focus only on the BPZ property and we do not require the derivation property. }
\begin{subequations}
\begin{align} 
\big{\langle } \, \mathbf{D'} \, A \, , \, B \, \big{\rangle }  = (-)^{\mathbf{D'} A} \big{\langle } \, A \, , \, \mathbf{D'} \,  B \, \big{\rangle } , 
\end{align} 
and for $\alpha = \eta ,\teta $, we can use the following cyclic and symmetric properties: 
\begin{align}
\big{\langle } \, A \, , \, \big[ \, B \, ,\,  C \, \big]^{\alpha }_{\Psi _{\eta \teta } }  \big{\rangle } 
= (-)^{AB} \big{\langle } \, B \, , \, \big[ \, A \, , \, C \, \big] ^{\alpha }_{\Psi _{\eta \teta }} \big{\rangle }
= (-)^{A(B+C)} \big{\langle } \, B \, , \, \big[ \, C \, , \, A \, \big] ^{\alpha }_{\Psi _{\eta \teta }} \big{\rangle } \, . 
\end{align}
For $\mathbf{D} = \partial _{t}$, $\delta $, or $Q$, because of the derivation properties of ${\bf L}^{\boldsymbol{\alpha }}$, we find 
\begin{align}
(-)^{\mathbf{D} } \mathbf{D} \, \big( \, D_{\alpha } \, A \, \big) - D_{\alpha } \big( \, \mathbf{D} \, A \, \big) - \big[ \, \mathbf{D} \, \Psi _{\eta \teta } \, , A \, \big] ^{\alpha }_{\Psi _{\eta \teta }} & = 0 \, , \hspace{5mm} (\alpha = \eta , \teta ) . 
\end{align}
\end{subequations}
In particular, note that with setting $A = \Psi _{t}$ and $B = D_{\eta } \Psi _{\delta }$, the relation (\ref{matching}) provides 
\begin{align}
\label{formula}
D_{\teta } \Big( D_{\eta } [ A , B ]^{\teta }_{\Psi _{\eta \teta }} 
+ [ D_{\teta } A , B ]^{\eta }_{\Psi _{\eta \teta }} 
+ [ D_{\eta } A , B ]^{\teta }_{\Psi _{\eta \teta }} 
+ [ A , D_{\teta } B ]^{\eta }_{\Psi _{\eta \teta }} \Big) = 0 . 
\end{align}

We prove that when we have WZW-like functional fields $\Psi _{\eta \teta } [ \varphi ]$ and $\Psi _{\mathbf{D} } [\varphi ]$ which satisfy (\ref{WZW 2}), our NS-NS action $S_{\eta \teta } [\varphi ]$ has topological $t$-dependence of (\ref{topo}). 
We carry out a direct computation of the variation of the action: 
\begin{align*}
\delta S_{\eta \tilde{\eta }} [\varphi ] = \int _{0}^{1} dt \Big( \big{\langle } \delta \Psi _{t} [\varphi (t) ] , \, Q \, \Psi _{\eta \tilde{\eta } } [\varphi (t) ] \big{\rangle } + \big{\langle } \Psi _{t} [\varphi (t) ] , \, \delta \big( Q \, \Psi _{\eta \tilde{\eta } } [\varphi (t) ] \big) \big{\rangle } \Big) . 
\end{align*}
For brevity, we omit $\varphi (t)$-dependence of functionals: We do not need it in computations. 
Using (3.8) in addition to (3.2) and (\ref{WZW 2}), 
we find that the second term can be transformed into $\langle \Psi _{\delta } , \partial _{t} (Q \Psi _{\eta \teta }) \rangle $ plus extra terms: 
\begin{subequations} 
\begin{align}
\big{\langle } \Psi _{t} , \, \delta ( Q \, \Psi _{\eta \teta } ) \big{\rangle } 
& = \langle \Psi _{t} ,  Q \, D_{\teta } D_{\eta } \Psi _{\delta }  \rangle 
\no 
& = \langle \Psi _{t} , D_{\teta } D_{\eta } Q  \Psi _{\delta } \rangle 
- \langle \Psi _{t} , [ Q \Psi _{\eta \teta } , \, D_{\eta } \Psi _{\delta } ]^{\teta }_{\Psi _{\eta \teta }} \rangle 
+ \langle \Psi _{t} , D_{\teta } [ Q \Psi _{\eta \teta } , \, \Psi _{\delta } ]^{\eta }_{\Psi _{\eta \teta }} \rangle
\no 
& = - \langle \Psi _{\delta } , Q D_{\eta } D_{\teta } \Psi _{t} \rangle 
- \langle Q \Psi _{\eta \teta } , [ \Psi _{t} , \, D_{\eta } \Psi _{\delta } ]^{\teta }_{\Psi _{\eta \teta }} \rangle 
- \langle Q \Psi _{\eta \teta } , [ D_{\teta } \Psi _{t} , \, \Psi _{\delta } ]^{\eta }_{\Psi _{\eta \teta }} \rangle 
\no 
& = \langle \Psi _{\delta } , \partial _{t} \big( Q \Psi _{\eta \teta } \big) \rangle 
+ \langle \Psi _{Q} , D_{\teta } D_{\eta } \Big( [ \Psi _{t} , \, D_{\eta } \Psi _{\delta } ]^{\teta }_{\Psi _{\eta \teta }} 
+ [ D_{\teta } \Psi _{t} , \, \Psi _{\delta } ]^{\eta }_{\Psi _{\eta \teta }} \Big) \rangle 
\no 
& = \big{\langle } \Psi _{\delta } , \, \partial _{t} \big( Q \Psi _{\eta \teta } \big) \big{\rangle } 
+ \big{\langle } \Psi _{Q} , \, [ D_{\teta } D_{\eta }\Psi _{t} , \, \Psi _{\delta } ]^{\eta }_{\Psi _{\eta \teta }} \rangle
\no & \hspace{15mm} 
+ \langle \Psi _{Q} , D_{\teta } \Big( D_{\eta } [ \Psi _{t} , \, D_{\eta } \Psi _{\delta } ]^{\teta }_{\Psi _{\eta \teta }} 
+ [ D_{\teta } \Psi _{t} , \, D_{\eta } \Psi _{\delta } ]^{\eta }_{\Psi _{\eta \teta }} \Big) \rangle . \label{2nd term}
\end{align}
Likewise, we find the first term of the variation becomes $\langle \partial _{t} \Psi _{\delta } , Q \Psi _{\eta \teta } \rangle $ plus extra terms: 
\begin{align}
\big{\langle } \delta \Psi _{t} , \, Q \, \Psi _{\eta \teta } \big{\rangle } 
& = - \langle D_{\teta } D_{\eta } \delta \Psi _{t} , \Psi _{Q} \rangle 
\no
& = - \langle \delta \big( D_{\teta } D_{\eta } \Psi _{t} \big) , \Psi _{Q} \rangle 
+ \langle [ \delta \Psi _{\eta \teta } , D_{\eta } \Psi _{t} ]^{\teta }_{\Psi _{\eta \teta }} , \Psi _{Q} \rangle 
+ \langle D_{\teta } [ \delta \Psi _{\eta \teta } , \Psi _{t} ]^{\eta }_{\Psi _{\eta \teta }} , \Psi _{Q} \rangle 
\no
& = - \langle \partial _{t} \big( \delta \Psi _{\eta \teta } \big) , \Psi _{Q} \rangle 
+ \langle \Psi _{Q} , \, [ D_{\eta } \Psi _{t} , \delta \Psi _{\eta \teta } ]^{\teta }_{\Psi _{\eta \teta }} 
+ D_{\teta } [  \Psi _{t} , \delta \Psi _{\eta \teta } ]^{\eta }_{\Psi _{\eta \teta }}  \rangle 
\no
& = - \langle \partial _{t} \big( D_{\teta } D_{\eta } \Psi _{\delta } \big) , \Psi _{Q} \rangle 
+ \langle \Psi _{Q} , \, [ D_{\eta } \Psi _{t} , D_{\teta } D_{\eta } \Psi _{\delta } ]^{\teta }_{\Psi _{\eta \teta }}
+ D_{\teta } [ \Psi _{t} , D_{\teta } D_{\eta } \Psi _{\delta } ]^{\eta }_{\Psi _{\eta \teta }}  \rangle 
\no
& = - \langle D_{\teta } D_{\eta } \partial _{t} \Psi _{\delta } , \Psi _{Q} \rangle 
- \langle [ \partial _{t} \Psi _{\eta \teta } , D_{\eta } \Psi _{\delta } ]^{\teta }_{\Psi _{\eta \teta } } 
+ D_{\teta } [ \delta \Psi _{\eta \teta } , \Psi _{\delta } ]^{\eta }_{\Psi _{\eta \teta } } , \Psi _{Q} \rangle 
\no & \hspace{10mm} 
+ \langle \Psi _{Q} , \, [ D_{\eta } \Psi _{t} , D_{\teta } D_{\eta } \Psi _{\delta } ]^{\teta }_{\Psi _{\eta \teta }}
+ D_{\teta } [ \Psi _{t} , D_{\teta } D_{\eta } \Psi _{\delta } ]^{\eta }_{\Psi _{\eta \teta }}  \rangle 
\no
& = \langle \partial _{t} \Psi _{\delta } , D_{\eta } D_{\teta } \Psi _{Q} \rangle 
- \langle \Psi _{Q} , \, [ D_{\teta } D_{\eta } \Psi _{t} , D_{\eta } \Psi _{\delta } ]^{\teta }_{\Psi _{\eta \teta } } 
+ D_{\teta } [ D_{\teta } D_{\eta } \Psi _{t} , \Psi _{\delta } ]^{\eta }_{\Psi _{\eta \teta } } \rangle 
\no & \hspace{10mm} 
+ \langle \Psi _{Q} , \, [ D_{\eta } \Psi _{t} , D_{\teta } D_{\eta } \Psi _{\delta } ]^{\teta }_{\Psi _{\eta \teta }} 
+ D_{\teta } [ \Psi _{t} , D_{\teta } D_{\eta } \Psi _{\delta } ]^{\eta }_{\Psi _{\eta \teta }}  \rangle 
\no
& = \big{\langle } \partial _{t} \Psi _{\delta } , \, Q \, \Psi _{\eta \teta } \big{\rangle } 
+ \big{\langle } \Psi _{Q} , \, [ D_{\eta } D_{\teta } \Psi _{t} , \Psi _{\delta } ]^{\eta }_{\Psi _{\eta \teta } }  \big{\rangle }
\no & \hspace{15mm} 
+ \big{\langle } \Psi _{Q} , \, D_{\teta } \Big( [ D_{\eta } \Psi _{t} , D_{\eta } \Psi _{\delta } ]^{\teta }_{\Psi _{\eta \teta } } 
+ [ \Psi _{t} , D_{\teta } D_{\eta } \Psi _{\delta } ]^{\eta }_{\Psi _{\eta \teta }} \Big) \big{\rangle } . \label{1st term}
\end{align}
\end{subequations} 
If and only if the sum of these extra terms vanishes, the action (\ref{Action}) has a topological $t$-dependence. 
However, (\ref{formula}) ensure the cancellation of these extra terms, and we find 
\begin{align*}
(\ref{2nd term}) + (\ref{1st term}) = \big{\langle } \partial _{t} \Psi _{\delta } , \, Q \, \Psi _{\eta \teta } \big{\rangle } + \big{\langle } \Psi _{\delta } , \, \partial _{t} \big( Q \, \Psi _{\eta \teta } \big) \big{\rangle } .
\end{align*} 
Using $\varphi (0) = 0$ and $\varphi (1) = \varphi$, it concludes our proof of (\ref{topo}): 
\begin{align*}
\delta S_{\eta \tilde{\eta }} [\varphi ] = \int _{0}^{1} dt \, \frac{\partial }{\partial t} \big{\langle } \Psi _{\delta } [\varphi (t) ] , \, Q \, \Psi _{\eta \tilde{\eta } } [\varphi (t) ] \big{\rangle } = \big{\langle } \Psi _{\delta } [\varphi ] , \, Q \, \Psi _{\eta \tilde{\eta } } [\varphi ] \big{\rangle } . 
\end{align*}

In summary, for fixed $L_{\infty }$ triplet $({\bf L}^{\boldsymbol{\eta }} , {\bf L}^{\boldsymbol{\teta } } \, ; {\bf Q})$, we first consider a functional $\Psi _{\eta \teta }$ satisfying constraint equations (\ref{WZW 1a}) and (\ref{WZW 1b}) defined by two of it, ${\bf L}^{\boldsymbol{\eta }}$ and ${\bf L}^{\boldsymbol{\teta }}$. 
Next, using this $\Psi _{\eta \teta }$, we estimate the WZW-like relation (\ref{WZW 2}) and derive the other functional $\Psi _{\mathbf{D}}$, which gives a half input of the action. 
Lastly, using $\Psi _{\eta \teta }$, we consider the Maurer-Cartan element of the remaining $L_{\infty }$ product ${\bf Q}$, which provides the on-shell condition (\ref{MC Q}) and thus the other half of the action. 
Then, combining these, we can obtain a gauge invariant WZW-like action (\ref{Action}). 

%\clearpage 

\section{Two constructions} 

As we showed in section 3, when two states $\Psi _{\eta \eta }[\varphi ]$ and $\Psi _{\mathbf{D}}[\varphi ]$ satisfying (\ref{WZW 2}) are obtained, one can find the WZW-like action (\ref{Action}). 
Therefore, the construction of actions is equivalent to finding explicit expressions of these functionals in terms of the dynamical string field $\varphi $. 

In this section, we present two different expressions of these $\Psi _{\eta \teta }$, $\Psi _{\mathbf{D}}$ using two different dynamical string fields $\Phi$ and $\Psi $. 
It gives two different realisations of our WZW-like action, which we call small-space parametrisation $S_{\eta \teta }[\Phi ]$ and large-space parametrisation $S_{\eta \teta }[\Psi ]$. 

Through these constructions, we also see that once we have $\Psi _{\eta \teta } [\varphi ]$ explicitly as a functional of $\varphi$, the other functional $\Psi _{\mathbf{D}} [\varphi ]$ can be {\it derived} from $\Psi _{\eta \teta } [\varphi ]$. 
It would suggest that $\Psi _{\eta \teta }$ is the fundamental ingredient in WZW-like theory, which we will discuss in the next section. 

\subsection*{Small-space parametrisation: $\varphi = \Phi $}

We write $\Phi $ for a NS-NS dynamical string field belonging to the small Hilbert space: $\eta \Phi = 0$ and $\teta \Phi = 0$. 
This $\Phi $ is a Grassmann even, total ghost number $2$, left-moving picture number $-1$, and right-moving picture number $-1$ state.

\niu{Pure-gauge-like (functional) field $\Psi _{\eta \teta } [ \Phi ]$} 

As a functional of $\Phi $, the pure-gauge-like field $\Psi _{\eta \widetilde{\eta } } = \Psi _{\eta \teta } [\Phi ]$ can be constructed by 
\begin{align}
\label{pure small}
\Psi _{\eta \widetilde{\eta }} [\Phi ] \equiv \pi _{1} \widehat{\bf G} \big( e^{\wedge \Phi } \big) \, . 
\end{align} 
Note that co-homomorphism $\widehat{\bf G}$ preserves the total ghost, left-moving picture, and right-moving picture numbers and this $\Psi _{\eta \teta } [\Phi ]$ has correct quantum numbers as a pure-gauge-like field. 
Thus, to show it, we have to check that (\ref{pure small}) indeed satisfies the constraint equations (\ref{WZW 1a}) and (\ref{WZW 1b}). 

Recall that in coalgebraic notation, we can write (\ref{WZW 1a}) and (\ref{WZW 1b}) as follows: 
\begin{align*}
\pi _{1} \, {\bf L}^{\boldsymbol \alpha } \big( e^{\wedge \Psi _{\eta \teta } [\varphi ]} \big) = 0 \, , \hspace{5mm}  (\alpha = \eta , \teta ) .  
\end{align*}
Since $\Psi _{\eta \teta } [\Phi ]$ is given by using the group-like element, the following relation holds: 
\begin{align*}
e^{\wedge \Psi _{\eta \teta } [\Phi ]} = e^{\wedge \pi _{1} \widehat{\bf G} ( e^{\wedge\Phi } ) }  = \widehat{\bf G} \big( e^{\wedge\Phi } \big) . 
\end{align*}
Because of (\ref{dual left}) and (\ref{dual right}), one can quickly find that (\ref{pure small}) satisfies 
\begin{align*}
{\bf L}^{\boldsymbol \alpha } \big(e^{\wedge \Psi _{\eta \teta }[\Phi ]} \big)  = ( \widehat{\bf G} \, {\boldsymbol \alpha } \, \widehat{\bf G}^{-1} ) \, \widehat{\bf G} \big( e^{\wedge \Phi }   \big) = \widehat{\bf G} \, {\boldsymbol \alpha } \big( e^{\wedge \Phi } \big) = 0 \, , \hspace{5mm} ( {\boldsymbol \alpha } = {\boldsymbol \eta } , {\boldsymbol \teta } ) \, , 
\end{align*}
which provides a proof that (\ref{pure small}) gives a pure-gauge-like (functional) field. 
In the last equality, we used the properties of the dynamical string fields: $\eta \, \Phi = 0$ and $\teta \, \Phi = 0$. 
Thus, in this small-space parametrisation $\varphi = \Phi$, it is the origin of all algebraic relations of WZW-like theory.

\niu{Associated (functional) field $\Psi _{\mathbf{D}} [\Phi ]$} 

Similarly, as functionals of $\Phi$, the associated (functional) field $\Psi _{\mathbf{D}} = \Psi _{\mathbf{D}} [\Phi ]$ with $\mathbf{D} = \partial _{t}$ or $\mathbf{D} = \delta $ can be constructed by 
\begin{subequations} 
\begin{align} 
\label{d small1}
\Psi _{\mathbf{D} } [\Phi ] \equiv \pi _{1} \widehat{\bf G} \big( {\xi \, \widetilde{\xi } \, \mathbf{D} \, \Phi } \wedge e^{\wedge \Phi } \big) \, ,  
\end{align} 
and the associated (associated) field $\Psi _{Q} [\Phi ]$ can be given by 
\begin{align}
\label{d small2}
\Psi _{Q} [\Phi ] \equiv \pi _{1} \widehat{\bf G} \, \boldsymbol{Q}_{\boldsymbol{\xi } \boldsymbol{\tilde{\xi }} } \big( e^{\wedge \Phi } \big) \, ,
\end{align}
\end{subequations}
where $\boldsymbol{Q}_{\boldsymbol{\xi } \boldsymbol{\tilde{\xi }} }$ is a coderivation operation which we will define below. 

\vspace{2mm} 

Recall that $\Psi _{\eta \teta }$ satisfies the constraint equations (\ref{WZW 1a}) and (\ref{WZW 1b}), and thus $\mathbf{D} \, \Psi _{\eta \eta } = D_{\eta}$-exact $= D_{\teta }$-exact holds, which implies the existence of $\Psi _{\mathbf{D}}$ satisfying (\ref{WZW 2}). 
One can derive an explicit form of the functional $\Psi _{\mathbf{D}} [\Phi ]$ from $\Psi _{\eta \teta } [\Phi ]$ in this manner. 

\vspace{2mm} 

Using the graded commutator of two coderivations $\boldsymbol{D}_{1}$ and $\boldsymbol{D}_{2}$, 
\begin{align*}
\Ld \, \boldsymbol{D}_{1} \, , \, \boldsymbol{D}_{2} \, \Rd \equiv \, \boldsymbol{D}_{1} \,\, \boldsymbol{D}_{2} -(-)^{\boldsymbol{D}_{1} \boldsymbol{D}_{2} } \boldsymbol{D}_{2} \,\, \boldsymbol{D}_{1} \, , 
\end{align*}
we can write $\ld \, {\bf L}^{\boldsymbol{\alpha }} , \boldsymbol{D} \, \rd = 0$ for the mutual commutative properties of ${\bf L}^{\boldsymbol{\alpha }} = \widehat{\bf G} \, \boldsymbol{\alpha } \, \widehat{\bf G}^{-1}$ for $\boldsymbol{\alpha } = \boldsymbol{\eta } , \boldsymbol{\teta }$\,. 
Note that Iff $\boldsymbol{D}$ is linear, the mutual commutativity $\ld \, {\bf L}^{\boldsymbol{\alpha }} , \boldsymbol{D} \, \rd = 0$ gives just the $\boldsymbol{D}$-derivation property. 
Then, we notice the following correspondence of the commutativity:  
\begin{align*}
\Ld \, \widehat{\bf G} \, \boldsymbol{\alpha } \, \widehat{\bf G}^{-1} , \, \boldsymbol{D} \, \Rd = 0 
\hspace{5mm} \iff \hspace{5mm} 
\Ld \, \boldsymbol{\alpha } \, ,  \widehat{\bf G}^{-1} \, \boldsymbol{D} \, \widehat{\bf G} \, \Rd = 0 \, .
\end{align*}
Namely, the co-derivation $\widehat{\bf G}^{-1} \, \boldsymbol{D} \, \widehat{\bf G}$ commutes with both $\boldsymbol{\eta }$ and $\boldsymbol{\teta }$\,. 
Hence, because of $\eta $-exactness and $\teta $-exactness, there exist a coderivation $\boldsymbol{D}_{\boldsymbol{\xi \tilde{\xi } }}$ such that 
\begin{align*}
\widehat{\bf G}^{-1} \, \boldsymbol{D} \, \widehat{\bf G} = - (-)^{\boldsymbol{D}} \Ld \, {\boldsymbol \eta } \, , \, \ld \, \boldsymbol {\teta } \, , \, \boldsymbol{D}_{\boldsymbol{\xi \tilde{\xi } }} \, \rd \, \Rd \, . 
\end{align*}
Note that any derivation $\mathbf{D}$ can be lift to the corresponding coderivation, for which we also write $\mathbf{D}$, because it is a linear map. 
For example, when $\mathbf{D} = \partial _{t}$ and $\mathbf{D} = \delta$, the above coderivation $\boldsymbol{D}_{\boldsymbol{\xi \tilde{\xi } }}$ is just the operations assigning $\xi \tilde{\xi } \, \mathbf{D}$ on each slot because of $\mathbf{D}\, \widehat{\bf G} = \widehat{\bf G} \, \mathbf{D}$\,. 
Using $\boldsymbol{D}_{\boldsymbol{\xi } \boldsymbol{\tilde{\xi } }}$ and the properties of the dynamical string field, $\eta \, \Phi = 0$ and $\teta \, \Phi = 0$, we find 
\begin{align*}
(-)^{\boldsymbol{D}} \boldsymbol{D} \, \widehat{\bf G} \, \big( e^{\wedge \Phi } \big) 
& = - \widehat{\bf G} \, \Ld \, {\boldsymbol \eta } \, , \, \ld \, \boldsymbol {\teta } \, , \, \boldsymbol{D}_{\boldsymbol{\xi \tilde{\xi } }} \, \rd \, \Rd \, \big( e^{\wedge \Phi } \big) 
= - \widehat{\bf G} \, {\boldsymbol \eta } \, \boldsymbol{\teta } \, \boldsymbol{D}_{\boldsymbol{\xi } \boldsymbol{\tilde{\xi }} } \big( e^{\wedge \Phi } \big) 
\no  & 
= - {\bf L}^{\boldsymbol \eta } \, {\bf L}^{\boldsymbol \teta } \, \widehat{\bf G} \, \boldsymbol{D}_{\boldsymbol{\xi } \boldsymbol{\tilde{\xi }} } \, \big( e^{\wedge \Phi } \big) 
\no 
& = - {\bf L}^{\boldsymbol \eta } \Big(  
\pi _{1} {\bf L}^{\boldsymbol \teta } \, \big( 
\pi _{1} \widehat{\bf G} \, \boldsymbol{D}_{\boldsymbol{\xi } \boldsymbol{\tilde{\xi }} } \, ( e^{\wedge \Phi } ) 
\wedge e^{\wedge\pi _{1} \widehat{\bf G}(e^{\wedge \Phi }) } \big) 
\wedge e^{\wedge \pi _{1} \widehat{\bf G} (e^{\wedge \Phi }) } \Big) \, . 
\end{align*} 
Note that with (\ref{pure small}), the linear operator $D_{\alpha }$ for $\alpha = \eta , \widetilde{\eta }$ can be written as 
\begin{align*}
D_{\alpha } = \pi _{1} {\bf L}^{\boldsymbol \alpha } \big( \, \mathbb{I} \wedge e^{\wedge \pi _{1} \widehat{\bf G} ( e^{\wedge \Phi  })} \big) , \hspace{5mm} ( {\boldsymbol \alpha } = {\boldsymbol \eta } , \widetilde{\boldsymbol \eta } ) \, . 
\end{align*}
We thus find that if we define the associated field $\Psi _{\mathbf{D}} [\Phi ]$ by the following functional of $\Phi $, 
\begin{align*} 
\Psi _{\mathbf{D}} [\Phi ] \equiv \pi _{1} \widehat{\bf G} \, \boldsymbol{D}_{\boldsymbol{\xi } \boldsymbol{\tilde{\xi }} } \, ( e^{\wedge \Phi } ) \, , 
\end{align*}
which reduces to (\ref{d small1}) and (\ref{d small2}), the Wess-Zumino-Witten-like relation (\ref{WZW 2}) indeed holds: 
\begin{align*}
(-)^{\mathbf{D}} \mathbf{D} \, \Psi _{\eta \widetilde{\eta }} [ \Phi ] = - D_{\eta } \, D_{\teta } \, \Psi _{\mathbf{D}} [\Phi ] . 
\end{align*}

\subsection*{Large-space parametrisation: $\varphi = \Psi $} 

We write $\Psi $ for a dynamical NS-NS string field which belongs to the left-and-right large Hilbert space: $\eta \Psi \not= 0$, $\teta \Psi  \not= 0$, and $\eta \teta \Psi \not= 0$. 
This $\Psi $ has total ghost number $0$, left-moving picture number $0$, and right-moving picture number $0$. 

\niu{Pure-gauge-like (functional) field $\Psi _{\eta \teta } [\Psi ]$} 

Let us consider the solution $\Psi _{\eta \teta } [ \tau ; \Psi ]$ of the following differential equation, 
\begin{align}
\label{pure eq}
\frac{\partial }{\partial \tau } \Psi_{\eta \teta } [\tau ; \Psi  ] = D_{\eta } (\tau ) D_{\teta } (\tau ) \, \Psi  
\end{align}
with the initial condition $\Psi _{\eta \teta } [\tau = 0 ; \Psi ] = 0$, where for any state $A \in \mathcal{H}$, we define 
\begin{align*}
D_{\alpha }(\tau ) A \equiv \alpha \, A + \sum_{n=0}^{\infty } \frac{1}{n!} \big[ \overbrace{\Psi _{\eta \teta } [ \tau ; \Psi ] , \dots , \Psi _{\eta \teta } [ \tau ; \Psi ] }^{n}, A \big] ^{\alpha } \, , 
\hspace{3mm} (\alpha = \eta , \teta ) \, . 
\end{align*} 
A pure-gauge-like (functional) field $\Psi _{\eta \teta }[ \Psi  ]$ is obtained as the $\tau = 1$ value solution 
\begin{align}
\label{pure large}
\Psi _{\eta \teta } [\Psi ] \equiv \Psi _{\eta \teta } [\tau =1 ; \Psi ] . 
\end{align} 
Note that (\ref{pure eq}) has the same form as the defining equation of a pure gauge field in bosonic string field theory \cite{Schubert:1991en}, which is the origin of the name {\it pure-gauge-like (functional) field}. 
We check that this $\Psi _{\eta \teta } [ \Psi ]$ satisfies (\ref{WZW 1a}) and (\ref{WZW 1b}). 
For this purpose, we set 
\begin{align*}
\mathcal{MC}_{L^{\alpha }}( \tau ) \equiv \, \alpha \, \Psi _{\eta \widetilde{\eta }} [ \tau ; \Psi ] + \sum_{n=1}^{\infty } \frac{1 }{n!} \big[ \overbrace{ \Psi _{\eta \widetilde{\eta }} [ \tau ; \Psi ] \, , \dots , \Psi _{\eta \widetilde{\eta }} [\tau ; \Psi ] }^{n} \big] ^{\alpha }  \, , 
\hspace{5mm} (\alpha = \eta , \teta ) . 
\end{align*}
Because of the initial condition $\Psi _{\eta \teta } [0 ; \Psi ] = 0$ of (\ref{pure eq}), it satisfies $\mathcal{MC}_{L^{\alpha }} ( 0 ) = 0$. 
Using (\ref{pure eq}) and (\ref{weak L1}), we obtain the following linear differential equation 
\begin{align}
\label{pure proof}
\frac{\partial }{\partial \tau } \mathcal{MC}_{L^{\alpha }} (\tau ) & = D_{\alpha } ( \tau ) \, \partial _{\tau } \Psi _{\eta \teta } [ \tau ; \Psi ] 
\no 
&= ( - )^{|\alpha |} \big[ \, \mathcal{MC}_{L^{\alpha }} (\tau ) \, , D_{\tilde{\alpha } } (\tau ) \Psi \, \big] ^{\alpha }_{\Psi _{\eta \teta } [\tau ; \Psi ]} \, ,
\end{align}
where $(-)^{|\alpha |}$ denotes $-1$ for $\alpha = \eta $ and $+1$ for $\alpha = \teta $. 
The initial condition $\mathcal{MC}_{L^{\alpha }} (\tau )=0$ provides that we have $\mathcal{MC}_{L^{\alpha }} (\tau ) = 0$ for any $\tau $, which ensures (\ref{pure large}) indeed satisfies (\ref{WZW 1a}) and (\ref{WZW 1b}) and gives a proof that (\ref{pure large}) is a pure-gauge-like (functional) field. 
By the iterated integral of (\ref{pure eq}), one can quickly find that a few terms of (\ref{pure large}) are given by
\begin{align*}
\Psi _{\eta \teta } [ \Psi ] = \eta \teta \, \Psi + \frac{1}{2} \Big( \big[ \eta \teta \, \Psi , \teta \, \Psi \big] ^{\eta } + \eta \, \big[ \eta \teta \, \Psi , \Psi \big] ^{\teta } \Big) + \dots  \, . 
\end{align*}
In this parametrisation, the properties of the dynamical string field $\eta \, \teta \, \Psi \not= 0$ makes possible to use $\Psi $ itself just like a gauge parameter of the nilpotent transformations generated by ${\bf L}^{\boldsymbol{\teta }}$ and ${\bf L}^{\boldsymbol{\teta }}$, and to have a pure-gauge-like field $\Psi _{\eta \teta }[\Psi ]$ as a functional of $\Psi $. 
(Note that they are not the gauge transformations of our theory; it only reminds us those of other theories.)

\niu{Associated (functional) field $\Psi _{\mathbf{D}} [ \Psi ]$}

We consider the following differential equation 
\begin{align}
\label{def1 of Psi_d}
- \frac{\partial }{\partial \tau } \Psi _{\mathbf{D} } [ \tau ; \Psi ] =  (-)^{\mathbf{D}} \mathbf{D} \, \Psi + \big[ D_{\eta } (\tau ) \Psi \, , \Psi _{\mathbf{D}} [ \tau ; \Psi ] \big] ^{\teta }_{\Psi _{\eta \teta } [\tau ; \Psi ] } + \big[ \Psi \, , D_{\teta } (\tau ) \Psi _{\mathbf{D}} [\tau ; \Psi ] \big] ^{\eta }_{\Psi _{\eta \teta } [\tau ; \Psi ] }  , 
\end{align}
with the initial condition $\Psi _{\mathbf{D}} [0;\Psi ] = 0$ {\it up to $D_{\eta }$-exact or $D_{\teta }$-exact terms}. 
An associated (functional) field $\Psi _{\mathbf{D}} [\Psi ]$ is obtained by the $\tau = 1$ value solution of (\ref{def1 of Psi_d}), 
\begin{align}
\label{d large} 
\Psi _{\mathbf{D}} [ \Psi ] \equiv \Psi _{\mathbf{D}} [ \tau = 1 ; \Psi ] \, . 
\end{align}
As $D_{\eta }$-exacts and $D_{\teta }$-exacts does not affect in the first slot of (\ref{Action}), this $\Psi _{\mathbf{D}}$ is determined up to these. 
To prove (\ref{d large}) satisfy (\ref{WZW 2}), we set 
\begin{align*} 
\mathcal{I} (\tau ) \equiv D_{\eta } D_{\teta } \, \Psi _{\mathbf{D}} [\tau ; \Psi ] + (-)^{\mathbf{D}} \, \mathbf{D} \, \Psi _{\eta \teta } [ \tau ; \Psi ] . 
\end{align*}
Note that iff we prove $\mathcal{I} (\tau ) = 0$ for any $\tau $, it implies we have an appropriate associated field $\Psi _{\mathbf{D}} [\Psi ]$. 
Using (3.2) and (\ref{pure eq}), we find 
\begin{align}
\frac{\partial }{\partial \tau } \mathcal{I} (\tau ) 
& = \big[ \partial _{\tau } \Psi _{\eta \teta } , D_{\teta } \Psi _{\mathbf{D}} \big] ^{\eta }_{\Psi _{\eta \teta }} + D_{\eta } \big[ \partial _{\tau } \Psi _{\eta \teta } , \Psi _{\mathbf{D}} \big]^{\teta }_{\Psi _{\eta \teta }}
 + D_{\eta } D_{\teta } \, \partial _{\tau } \Psi _{\mathbf{D}} + (-)^{\mathbf{D}} \mathbf{D} \, \partial _{\tau } \Psi _{\eta \teta }   
\no 
& = \big[ D_{\eta } D_{\teta } \Psi , D_{\teta } \Psi _{\mathbf{D}} \big] ^{\eta }_{\Psi _{\eta \teta }} + D_{\eta } \big[ D_{\eta } D_{\teta } \Psi , \Psi _{\mathbf{D}} \big]^{\teta }_{\Psi _{\eta \teta }}
 + D_{\eta } D_{\teta } \, \partial _{\tau } \Psi _{\mathbf{D}} + (-)^{\mathbf{D}} \mathbf{D} \, D_{\eta } D_{\teta } \Psi  
\no 
& = \Big{\{ } - D_{\eta } \big[ D_{\teta } \Psi , D_{\teta } \Psi _{\mathbf{D}} \big] ^{\eta }_{\Psi _{\eta \teta }} 
+ \big[ D_{\teta } \Psi , D_{\eta } D_{\teta } \Psi _{\mathbf{D}} \big] ^{\eta }_{\Psi _{\eta \teta }} \Big{\} }
- D_{\eta } \big[ D_{\eta } \Psi , D_{\teta } \Psi _{\mathbf{D}} \big] ^{\teta }_{\Psi _{\eta \teta }}
\no & \hspace{25mm} 
+ \Big{\{ } \big[ \mathbf{D} \, \Psi _{\eta \teta } , D_{\teta } \Psi \big] ^{\eta }_{\Psi _{\eta \teta }} 
+ D_{\eta } \big[ (-)^{\mathbf{D}} \mathbf{D} \, \Psi _{\eta \teta } , \Psi \big] ^{\teta }_{\Psi _{\eta \teta } }  \Big{\} } 
\no & \hspace{10mm} 
+ D_{\eta } D_{\teta } \Big( \partial _{\tau } \Psi _{\mathbf{D}} + (-)^{\mathbf{D}} \mathbf{D} \, \Psi + \big[ D_{\eta } \Psi , \Psi _{\mathbf{D}} \big]^{\teta }_{\Psi _{\eta \teta }}  \Big) 
\no 
& = \big[ D_{\teta } \Psi , \mathcal{I} (\tau ) \big] ^{\eta }_{\Psi _{\eta \teta }} 
+ D_{\eta } \big[ \Psi , \mathcal{I}(\tau ) \big] ^{\teta }_{\Psi _{\eta \teta } }  
\no & \hspace{10mm} 
+ D_{\eta } D_{\teta } \Big( \partial _{\tau } \Psi _{\mathbf{D}} + (-)^{\mathbf{D}} \mathbf{D} \, \Psi + \big[ D_{\eta } \Psi , \Psi _{\mathbf{D}} \big]^{\teta }_{\Psi _{\eta \teta }} + \big[ \Psi , D_{\teta } \Psi _{\mathbf{D}} \big] ^{\eta }_{\Psi _{\eta \teta }} \Big) . \label{way1}
\end{align}
From the third equal to the forth equal, we used the following identity: 
\begin{align*}
- D_{\eta } \big[ D_{\eta } \Psi , D_{\teta } \Psi _{\mathbf{D}} \big] ^{\teta }_{\Psi _{\eta \teta }}
& = D_{\eta } D_{\teta } \big[ \Psi , D_{\teta } \Psi _{\mathbf{D}} \big] ^{\eta }_{\Psi _{\eta \teta }} 
+ D_{\eta } \big[ D_{\teta } \Psi , D_{\teta } \Psi _{\mathbf{D}} \big] ^{\eta }_{\Psi _{\eta \teta }}
+ D_{\eta } \big[ \Psi , ( D_{\teta } )^{2} \Psi _{\mathbf{D}} \big] ^{\eta }_{\Psi _{\eta \teta }}
\no & \hspace{10mm}
+ ( D_{\eta } )^{2} \big[ \Psi , D_{\teta } \Psi _{\mathbf{D}} \big] ^{\teta }_{\Psi _{\eta \teta }}
+ D_{\eta } \big[ \Psi , D_{\eta } D_{\teta } \Psi _{\mathbf{D}} \big] ^{\teta }_{\Psi _{\eta \teta }} 
\no 
& = D_{\eta } D_{\teta } \big[ \Psi , D_{\teta } \Psi _{\mathbf{D}} \big] ^{\eta }_{\Psi _{\eta \teta }} 
+ D_{\eta } \Big( \big[ D_{\teta } \Psi , D_{\teta } \Psi _{\mathbf{D}} \big] ^{\eta }_{\Psi _{\eta \teta }}
+ \big[ \Psi , D_{\eta } D_{\teta } \Psi _{\mathbf{D}} \big] ^{\teta }_{\Psi _{\eta \teta }} \Big) . 
\end{align*}
When $\Psi _{\mathbf{D}} [ \tau ; \Psi ]$ satisfies (\ref{def1 of Psi_d}) up to $D_{\eta }$ -exacts and $D_{\teta }$-exacts, we have 
\begin{align*} 
\frac{\partial }{\partial \tau } \mathcal{I} (\tau ) = \big{\{ } \Psi \, , \, \mathcal{I}(\tau ) \big{\} } _{\Psi _{\eta \teta } (\tau )} \, ,
\end{align*}
which is the same type of differential equations as (\ref{pure proof}), where $\{ A , B \} _{\Psi _{\eta \teta }}$ is defined by 
\begin{align*} 
\big{\{ } A \, , B \big{\} }_{\Psi _{\eta \teta } (\tau )} \equiv \big[ D_{\teta }(\tau ) A \, , B \big] ^{\eta }_{\Psi _{\eta \teta } [\tau ; \Psi ] }  + D_{\eta } (\tau ) \big[ A \, , B \big] ^{\teta }_{\Psi _{\eta \teta } [\tau ; \Psi ] }  \, . 
\end{align*}
The initial condition $\cI (0) = 0$ provides that we have $\cI ( \tau ) = 0$ for any $\tau $, which gives a proof that (\ref{d large}) satisfies (\ref{WZW 2}). 
For example, one can quickly find a few terms of $\Psi _{t} [\Psi (t) ]$ are 
\begin{align*}
\Psi _{t} [\Psi (t) ] = - \partial _{t} \Psi (t) + \frac{1}{2} \Big( \big[ \eta \, \Psi (t) , \partial _{t} \Psi (t) \big] ^{\teta } + \big[ \Psi (t) , \teta \, \partial _{t} \Psi (t) \big] ^{\eta } \Big) + \dots \, . 
\end{align*}

\niu{On the $D_{\eta }$-exacts and $D_{\teta }$-exacts}
 
We found a defining equation (\ref{def1 of Psi_d}) of $\Psi _{\mathbf{D}} [\Psi ]$. 
Since it is up to $D_{\eta }$-exacts and $D_{\teta }$-exacts, one can find another expression. 
Note that we have the following identity 
\begin{align*}
& \Big{( } \big[ D_{\eta } \Psi \, , \Psi _{\mathbf{D}} \big] ^{\teta }_{\Psi _{\eta \teta }} 
+ \big[  \Psi , D_{\teta } \Psi _{\mathbf{D}} \big] ^{\eta }_{\Psi _{\eta \teta }} \Big{) } 
+ 
\Big{(} \big[ D_{\teta } \Psi , \Psi _{\mathbf{D}} \big] ^{\eta }_{\Psi _{\eta \teta } }   
+ \big[ \Psi \, , D_{\eta } \Psi _{\mathbf{D}} \big] ^{\teta }_{\Psi _{\eta \teta }} \Big{)} 
\no & \hspace{30mm} = 
- D_{\eta } \big[ \Psi \, , \Psi _{\mathbf{D}} \big] ^{\teta }_{\Psi _{\eta \teta }}
- D_{\teta } \big[ \Psi \, , \Psi _{\mathbf{D}} \big] ^{\eta }_{\Psi _{\eta \teta }} 
\end{align*}
which provides another expression of (\ref{way1}): 
\begin{align*}
\frac{\partial }{\partial \tau } \mathcal{I} (\tau ) 
& = 
\big{\{ } \Psi \, , \, \mathcal{I}(\tau ) \big{\} } _{\Psi _{\eta \teta } (\tau )}
+ D_{\eta } D_{\teta } \Big( \partial _{\tau } \Psi _{\mathbf{D}} + (-)^{\mathbf{D}} \mathbf{D} \, \Psi 
- \big[ D_{\teta } \Psi , \Psi _{\mathbf{D}} \big]^{\eta }_{\Psi _{\eta \teta }} 
- \big[ \Psi , D_{\eta } \Psi _{\mathbf{D}} \big] ^{\teta }_{\Psi _{\eta \teta }} \Big) 
\end{align*}
It ensures that as a defining equation of $\Psi _{\mathbf{D}} [\tau ; \Psi ]$, we can also use 
\begin{align}
\partial _{\tau } \Psi _{\mathbf{D}} = - (-)^{\mathbf{D}} \mathbf{D} \, \Psi 
+ \big[ D_{\teta } \Psi , \Psi _{\mathbf{D}} \big]^{\eta }_{\Psi _{\eta \teta }} 
+ \big[ \Psi , D_{\eta } \Psi _{\mathbf{D}} \big] ^{\teta }_{\Psi _{\eta \teta }} . \label{def2 of Psi_d}
\end{align}
The difference between (\ref{def1 of Psi_d}) and (\ref{def2 of Psi_d}) is just $D_{\eta }$-exacts plus $D_{\teta }$-exacts, which does not affect WZW-like relations and the resultant action: It is just the gauge invariance generated by $D_{\eta }$ and $D_{\teta }$. 
Note also that since we have $D_{\eta } (\tau ) D_{\teta } (\tau ) = - D_{\teta } (\tau ) D_{\eta } (\tau )$, one may compute as  
\begin{align}
\frac{\partial }{\partial \tau } \mathcal{I} (\tau ) 
& = \partial _{\tau } \big( - D_{\teta } D_{\eta } \, \Psi _{\mathbf{D}} \big) + (-)^{\mathbf{D}} \mathbf{D} \, \partial _{\tau } \Psi _{\eta \teta }   
\no 
%& = - \big[ \partial _{\tau } \Psi _{\eta \teta } , D_{\eta } \Psi _{d} \big] ^{\teta }_{\Psi _{\eta \teta }} - D_{\teta } \big[ \partial _{\tau } \Psi _{\eta \teta } , \Psi _{d} \big]^{\eta }_{\Psi _{\eta \teta }}
% - D_{\teta } D_{\eta } \, \partial _{\tau } \Psi _{d} + (-)^{d} d \, \partial _{\tau } \Psi _{\eta \teta }   
%\no 
%& = \big[ D_{\teta } D_{\eta } \Psi , D_{\eta } \Psi _{d} \big] ^{\teta }_{\Psi _{\eta \teta }} + D_{\teta } \big[ D_{\teta } D_{\eta } \Psi , \Psi _{d} \big]^{\eta }_{\Psi _{\eta \teta }}
% + D_{\eta } D_{\teta } \, \partial _{\tau } \Psi _{d} + (-)^{d} d \, D_{\eta } D_{\teta } \Psi  
%\no 
%& = \Big{\{ } - D_{\teta } \big[ D_{\eta } \Psi , D_{\eta } \Psi _{d} \big] ^{\teta }_{\Psi _{\eta \teta }} 
%+ \big[ D_{\eta } \Psi , D_{\teta } D_{\eta } \Psi _{d} \big] ^{\teta }_{\Psi _{\eta \teta }} \Big{\} }
%- D_{\teta } \big[ D_{\teta } \Psi , D_{\eta } \Psi _{d} \big] ^{\eta }_{\Psi _{\eta \teta }}
%\no & \hspace{25mm} 
%- \Big{\{ } \big[ d \Psi _{\eta \teta } , D_{\eta } \Psi \big] ^{\teta }_{\Psi _{\eta \teta }} 
%+ D_{\teta } \big[ (-)^{d} d \Psi _{\eta \teta } , \Psi \big] ^{\eta }_{\Psi _{\eta \teta } }  \Big{\} } 
%\no & \hspace{10mm} 
%+ D_{\eta } D_{\teta } \Big( \partial _{\tau } \Psi _{d} + (-)^{d} d \Psi - \big[ D_{\teta } \Psi , \Psi _{d} \big]^{\eta }_{\Psi _{\eta \teta }}  \Big) 
%\no 
& = - \big[ D_{\eta } \Psi , \mathcal{I} (\tau ) \big] ^{\teta }_{\Psi _{\eta \teta }} 
- D_{\teta } \big[ \Psi , \mathcal{I}(\tau ) \big] ^{\eta }_{\Psi _{\eta \teta } }  
\no & \hspace{10mm} 
+ D_{\eta } D_{\teta } \Big( \partial _{\tau } \Psi _{\mathbf{D}} + (-)^{\mathbf{D}} \mathbf{D} \, \Psi - \big[ D_{\teta } \Psi , \Psi _{\mathbf{D}} \big]^{\eta }_{\Psi _{\eta \teta }} - \big[ \Psi , D_{\eta } \Psi _{\mathbf{D}} \big] ^{\teta }_{\Psi _{\eta \teta }} \Big) . \label{way2}
\end{align}
%Here, the following identity is usefull 
%\begin{align*}
%- D_{\teta } \big[ D_{\teta } \Psi , D_{\eta } \Psi _{d} \big] ^{\eta }_{\Psi _{\eta \teta }}
%& = D_{\teta } D_{\eta } \big[ \Psi , D_{\eta } \Psi _{d} \big] ^{\teta }_{\Psi _{\eta \teta }} 
%+ D_{\teta } \big[ D_{\eta } \Psi , D_{\eta } \Psi _{d} \big] ^{\teta }_{\Psi _{\eta \teta }}
%+ D_{\teta } \big[ \Psi , ( D_{\eta } )^{2} \Psi _{d} \big] ^{\teta }_{\Psi _{\eta \teta }}
%\no & \hspace{10mm}
%+ ( D_{\teta } )^{2} \big[ \Psi , D_{\eta } \Psi _{d} \big] ^{\eta }_{\Psi _{\eta \teta }}
%+ D_{\teta } \big[ \Psi , D_{\teta } D_{\eta } \Psi _{d} \big] ^{\eta }_{\Psi _{\eta \teta }} 
%\no 
%& = D_{\teta } D_{\eta } \big[ \Psi , D_{\eta } \Psi _{d} \big] ^{\teta }_{\Psi _{\eta \teta }} 
%+ D_{\teta } \Big( \big[ D_{\eta } \Psi , D_{\eta } \Psi _{d} \big] ^{\teta }_{\Psi _{\eta \teta }}
%+ \big[ \Psi , D_{\teta } D_{\eta } \Psi _{d} \big] ^{\eta }_{\Psi _{\eta \teta }} \Big) . 
%\end{align*}
However, we have the following identity 
\begin{align*}
& \Big( \big[ D_{\teta } \Psi , \mathcal{I} (\tau ) \big]^{\eta }_{\Psi _{\eta \teta }}
+ D_{\eta } \big[ \Psi , \mathcal{I} (\tau ) \big]^{\teta }_{\Psi _{\eta \teta }} \Big) 
+ \Big( \big[ D_{\eta } \Psi , \mathcal{I} (\tau ) \big]^{\teta }_{\Psi _{\eta \teta }} 
+ D_{\teta } \big[ \Psi , \mathcal{I} (\tau ) \big]^{\eta }_{\Psi _{\eta \teta }} \Big) 
\no & \hspace{30mm} 
= - \big[ \Psi , D_{\eta } \mathcal{I} (\tau ) \big]^{\teta }_{\Psi _{\eta \teta }}
- \big[ \Psi , D_{\teta } \mathcal{I} (\tau ) \big]^{\eta }_{\Psi _{\eta \teta }} . 
\end{align*}
Comparing (\ref{way1}) and (\ref{way2}) with (\ref{def2 of Psi_d}), we also find  
\begin{align*}
0 = \big[ \Psi , D_{\eta } \mathcal{I} (\tau ) \big]^{\teta }_{\Psi _{\eta \teta }}
+ \big[ \Psi , D_{\teta } \mathcal{I} (\tau ) \big]^{\eta }_{\Psi _{\eta \teta }}
= \big[ \Psi ,  (-)^{\mathbf{D}} D_{\eta } \mathbf{D} \, \Psi _{\eta \teta } \big]^{\teta }_{\Psi _{\eta \teta }}
+ \big[ \Psi , (-)^{\mathbf{D}} D_{\teta } \mathbf{D} \, \Psi _{\eta \teta } \big]^{\eta }_{\Psi _{\eta \teta }} \, . 
\end{align*}
These term can appear or vanish in computations of $\partial _{\tau } \cI (\tau ) = \{ \Psi , \cI (\tau )\} _{\Psi _{\eta \teta } (\tau ) }$. 

\subsection*{On the small associated fields} 

We constructed two functionals $\Psi _{\eta \teta }[ \varphi ]$ and $\Psi _{\mathbf{D}} [ \varphi ]$. 
It is sufficient to give a WZW-like action explicitly. 
However, one can consider small associated (functional) fields defined by 
\begin{align}
\label{small d}
\Psi _{\eta {\rm D}} [ \varphi ] \equiv D_{\eta } \, \Psi _{\mathbf{D}} [\varphi ] , \hspace{5mm} 
\Psi _{{\rm D} \teta } [\varphi ] \equiv D_{\teta } \, \Psi _{\mathbf{D}} [\varphi ] . 
\end{align}
The WZW-like relation (\ref{WZW 2}) provides that they satisfy the following relations 
\begin{align}
\label{WZW small} 
\mathcal{J}_{\eta } [\varphi ] \equiv D_{\teta } \Psi _{\eta {\rm D}} [\varphi ] - (-)^{\mathbf{D}} \mathbf{D} \, \Psi _{\eta \teta } [\varphi ] = 0,  
\hspace{5mm}  
\mathcal{J}_{\teta } [\varphi ] \equiv D_{\eta } \Psi _{{\rm D} \teta } [\varphi ] + (-)^{\mathbf{D}} \mathbf{D} \, \Psi _{\eta \teta } [\varphi] = 0 . 
\end{align}
One may prefer these because of the analogy with the NS sector. 
For example, using $(-)^{\mathbf{D}} \mathbf{D} \, \Psi _{\eta \teta } = D_{\eta } \Psi _{{\rm D} \teta } = D_{\teta } \Psi _{\eta {\rm D}}$ with derivations $\mathbf{D}_{1}$ and $\mathbf{D}_{2}$ satisfying $\mathbf{D}_{1} \mathbf{D}_{2} = (-)^{\mathbf{D}_{1} \mathbf{D}_{2} } \mathbf{D}_{2} \mathbf{D}_{1} $, one can find\footnote{They follow from direct computations 
\begin{align*} 
\mathbf{D}_{1} \mathbf{D}_{2} \Psi _{\eta \teta } & = (-)^{\mathbf{D}_{2}} \mathbf{D}_{1} \Big(  D_{\eta } \Psi _{{\rm D}_{2} \teta } \Big) = (-)^{\mathbf{D}_{1} + \mathbf{D}_{2}} \Big( D_{\eta } \, \mathbf{D}_{1} \Psi _{{\rm D}_{2} \teta } + [ \mathbf{D}_{1} \Psi _{\eta \teta } , \Psi _{{\rm D}_{2} \teta } ]^{\eta }_{\Psi _{\eta \teta } } \Big) 
\no 
& = (-)^{\mathbf{D}_{1} + \mathbf{D}_{2}} \Big( D_{\eta } \, \mathbf{D}_{1} \Psi _{{\rm D}_{2} \teta } + (-)^{\mathbf{D}_{1}} [ D_{\eta } \Psi _{{\rm D}_{1} \teta } , \Psi _{{\rm D}_{2} \teta } ]^{\eta }_{\Psi _{\eta \teta } } \Big) , 
\\ 
(-)^{\mathbf{D}_{1} \mathbf{D}_{2}}  \mathbf{D}_{2} \mathbf{D}_{1} \Psi _{\eta \teta } & = (-)^{\mathbf{D}_{1} + \mathbf{D}_{2} + \mathbf{D}_{1} \mathbf{D}_{2}} \Big( D_{\eta } \, \mathbf{D}_{2} \, \Psi _{{\rm D}_{1} \teta } + (-)^{\mathbf{D}_{2}} [ D_{\eta } \Psi _{{\rm D}_{2} \teta } , \Psi _{{\rm D}_{1} \teta } ]^{\eta }_{\Psi _{\eta \teta } } \Big) 
\no 
&  = (-)^{\mathbf{D}_{1} + \mathbf{D}_{2} + \mathbf{D}_{1} \mathbf{D}_{2}} \Big{\{ } D_{\eta } \Big( \mathbf{D}_{2} \Psi _{{\rm D}_{1} \teta } - (-)^{\mathbf{D}_{2}} [ \Psi _{{\rm D}_{2} \teta } , \Psi _{{\rm D}_{1} \teta } ]^{\eta }_{\Psi _{\eta \teta } } \Big) + [ \Psi _{{\rm D}_{2} \teta } , D_{\eta } \Psi _{{\rm D}_{1} \teta } ]^{\eta }_{\Psi _{\eta \teta }} \Big{\} } . 
\end{align*} 
} 
\begin{align*}
\mathbf{D}_{1} \Psi _{{\rm D}_{2} \teta } - (-)^{\mathbf{D}_{1} \mathbf{D}_{2}} \mathbf{D}_{2} \Psi _{{\rm D}_{1} \teta } - (-)^{\mathbf{D}_{1}} \big[ \Psi _{{\rm D}_{1} \teta } , \Psi _{{\rm D}_{2} \teta } \big] ^{\eta }_{\Psi _{\eta \teta }} = D_{\eta }\mathchar`-{\rm exact} , 
\no
\mathbf{D}_{1} \Psi _{\eta {\rm D}_{2}} - (-)^{\mathbf{D}_{1} \mathbf{D}_{2}} \mathbf{D}_{2} \Psi _{\eta {\rm D}_{1}} -(-)^{\mathbf{D}_{1}} \big[ \Psi _{\eta {\rm D}_{1}} , \Psi _{\eta {\rm D}_{2}} \big] ^{\teta }_{\Psi _{\eta \teta }} = D_{\teta }\mathchar`-{\rm exact} . 
\end{align*}
On the basis of these functionals and relations, one can obtain another check of the gauge invariance of the action. 
For details in this direction, see appendix E of \cite{Goto:2015pqv}. 
In the rest of this section, we explain how one can construct explicit forms of these as functionals of $\Phi $ or $\Psi$. 

\niu{Small-space parametrisation}

It is easy to obtain these in terms of $\Phi$ because the analogy with the NS sector exactly works. 
We find that small associated (functional) fields $\Psi _{{\rm D} \widetilde{\eta }}$ and $\Psi _{\eta {\rm D}}$ are given by 
\begin{align*}
\Psi _{{\rm D} \widetilde{\eta }} [\Phi ] \equiv \pi _{1} \widehat{\bf G} \big( {\boldsymbol D}_{\boldsymbol{\xi }} \, e^{\wedge \Phi } \big) ,
\hspace{5mm}  
\Psi _{\eta {\rm D}} [\Phi ] & = \pi _{1} \widehat{\bf G} \big( \boldsymbol{D}_{\boldsymbol{\tilde{\xi }} } \, e^{\wedge \Phi } \big) , 
\end{align*}
where we used coderivations ${\boldsymbol D}_{\boldsymbol{\xi }}$ and ${\boldsymbol D}_{\boldsymbol{\tilde{\xi }}}$ such that 
\begin{align*}
\widehat{\bf G}^{-1} \, \boldsymbol{D} \, \widehat{\bf G} = - (-)^{\boldsymbol{D}} \Ld \, {\boldsymbol \eta } \, , \, \boldsymbol{D}_{\boldsymbol{\xi }} \, \Rd \,  ,
\hspace{5mm} 
\widehat{\bf G}^{-1} \, \boldsymbol{D} \, \widehat{\bf G} = - (-)^{\boldsymbol{D}} \Ld \, \boldsymbol {\teta } \, , \, \boldsymbol{D}_{\boldsymbol{\tilde{\xi } }} \, \Rd \, . 
\end{align*} 
It is consistent with (\ref{small d}). 
Note that $\mathbf{D} \, \widehat{\bf G} = \widehat{\bf G} \, \mathbf{D}$ for $\mathbf{D} =\partial _{t} , \delta $, but ${\bf Q} \, \widehat{\bf G} = \widehat{\bf G} \, {\bf L}^{\rm NS,NS}$. 

\niu{Large-space parametrisation}

The situation becomes somewhat complicated in the large-space parametrisation. 
One can construct small associated (functional) fields $\Psi _{{\rm D} \teta } [\Psi ]$ and $\Psi _{\eta {\rm D}} [\Psi ]$ as the $\tau =1$ value solutions, 
\begin{align*}
\Psi _{{\rm D} \teta } [\Psi ] \equiv \Psi _{{\rm D} \teta } [ \tau =1 ; \Psi ] , \hspace{5mm} \Psi _{\eta {\rm D}} [ \Psi ] \equiv \Psi _{\eta {\rm D}} [ \tau =1 ; \Psi ] ,
\end{align*}
of the following differential equations 
\begin{align*}
\frac{\partial }{\partial \tau } \Psi _{{\rm D} \teta } [\tau ; \Psi ] & =  \mathbf{D} \, D_{\teta } (\tau ) \, \Psi + \big[ D_{\teta } (\tau ) \Psi , \Psi _{{\rm D} \teta } [ \tau ; \Psi ] \big] ^{\eta }_{\Psi _{\eta \teta }[\tau ; \Psi ] }  , 
\\ 
- \frac{\partial }{\partial \tau } \Psi _{\eta {\rm D}} [\tau ; \Psi ] & = \mathbf{D} \, D_{\eta } (\tau ) \, \Psi  + \big[ D_{\eta } (\tau ) \Psi , \Psi _{\eta {\rm D}} [ \tau ; \Psi ] \big] ^{\teta }_{\Psi _{\eta \teta }[\tau ; \Psi ] }  , 
\end{align*}
with the initial conditions $\Psi _{{\rm D} \teta } [ \tau = 0 ; \Psi ] = 0$ and $\Psi _{\eta {\rm D}} [ \tau = 0 ; \Psi ] = 0$. 
The minus sign of the second equation comes from the ordering of $D_{\eta }$ and $D_{\teta }$ in the definition of (\ref{pure eq}). 
One can also check these satisfy (\ref{WZW small}) using (\ref{pure eq}) in the same manner as the NS sector: 
The equation
\begin{align*} 
\partial _{\tau } \mathcal{J}_{\eta } & = 
[ D_{\eta } D_{\teta } \Psi , \Psi _{\eta {\rm D}} ]^{\teta } + D_{\teta } \partial _{\tau } \Psi _{\eta {\rm D}} - (-)^{\mathbf{D}} \mathbf{D} \, D_{\eta }D_{\teta } \Psi 
\no 
& = D_{\teta } \Big(  \partial _{\tau } \Psi _{\eta {\rm D}} + [D_{\eta } \Psi , \Psi _{\eta {\rm D}} ]^{\teta } + \mathbf{D} D_{\eta } \Psi \Big) - [ D_{\eta } \Psi , D_{\teta } \Psi _{\eta {\rm D}} - (-)^{\mathbf{D}} \mathbf{D} \, \Psi _{\eta \teta } ]^{\teta }
\end{align*}
with $\mathcal{J}_{\eta } (0) = 0$ provides $\mathcal{J}_{\eta } (\tau ) = 0$ for any $\tau $. 
Likewise, we find $\mathcal{J}_{\teta } (\tau ) = 0$ for any $\tau $. 

We can therefore obtain $\Psi _{\eta {\rm D}}$ and $\Psi _{{\rm D} \teta }$ satisfying (\ref{WZW small}) without using $\Psi _{\mathbf{D}}$ and (\ref{small d}). 
When we start with $\Psi _{\mathbf{D}}$ and (\ref{def1 of Psi_d}), does $D_{\eta } \Psi _{\mathbf{D}}$ or $D_{\eta } \Psi _{\mathbf{D}}$ of (\ref{small d}) satisfy the above differential equation? 
The answer is yes; it gives correct solutions up to $D_{\eta }$-exacts and $D_{\teta }$-exacts:  
\begin{align*} 
\frac{\partial }{\partial \tau } \big( D_{\eta }(\tau ) \Psi _{\mathbf{D}} [ \tau ; \Psi ] \big) % & = 
%D_{\eta } \Big( - (-)^{d} d \Psi - [ D_{\eta } \Psi , \Psi _{d} ]^{\teta }_{\Psi _{\eta \teta }} - [ \Psi , D_{\teta } \Psi _{d} ]^{\eta }_{\Psi _{\eta \teta }} \Big) + \big[ D_{\eta } D_{\teta } \Psi , \Psi _{d} \big] ^{\eta }_{\Psi _{\eta \teta }} 
%\no 
%& = - d ( D_{\eta } \Psi ) - [ d \Psi _{\eta \teta } , \Psi ]^{\eta } - D_{\eta } \big( [ \Psi , D_{\teta } \Psi _{d} ]^{\eta } + [ D_{\teta } \Psi , \Psi _{d} ]^{\eta } \big) 
%\no & \hspace{20mm} - D_{\eta } [ D_{\eta } \Psi , \Psi _{d} ]^{\teta } + [ D_{\teta } \Psi , D_{\eta } \Psi _{d} ]^{\eta }    
%\no & 
 = - \mathbf{D} \, ( D_{\eta } \, \Psi _{\mathbf{D}} ) - \big[ D_{\eta } \Psi \, , ( D_{\eta } \, \Psi _{\mathbf{D}} ) \big] ^{\teta }_{\Psi _{\eta \teta } } + D_{\teta } \big[ D_{\eta } \Psi \, , \Psi _{\mathbf{D}} \big] ^{\eta }_{\Psi _{\eta \teta } }  \, . 
\end{align*} 
%In the last equality, we used $- D_{\eta } [ D_{\eta } \Psi , \Psi _{d} ]^{\teta } = - [ D_{\eta } \Psi , D_{\eta } \Psi _{d} ]^{\teta } + D_{\teta } [ D_{\eta } \Psi , \Psi _{d} ]^{\eta } + [ D_{\eta } D_{\eta } \Psi , \Psi _{d} ]^{\eta } - [ D_{\eta } \Psi , D_{\teta } \Psi _{d } ] ^{\eta }$. 
Conversely, when we set $D_{\eta } A = \Psi _{\eta {\rm D}}[\Psi ]$ and start with these differential equations, can we derive the fact that this $A$ satisfies (\ref{def1 of Psi_d})? 
The answer is again yes; we can re-derive (\ref{def1 of Psi_d}) up to $D_{\eta }$-exacts and $D_{\teta }$-exacts. 
Thus large and small associated fields both work well. 

\vspace{2mm}

\niu{On the $D_{\eta }$-exactness and $D_{\teta }$-exactness}

We can only specify the large associated (functional) field $\Psi _{\mathbf{D}}$ up to $D_{\eta }$- and $D_{\teta }$-exact terms, and these ambiguities do not contribute in the action. 
Therefore, in principle, one could set these any values by hand. 
We have operators $F \xi $ and $\widetilde{F} \tilde{\xi }$ defined by 
\begin{align} 
\label{F}
F \xi \equiv \sum_{n=0}^{\infty } \Big[ \xi \big( \eta - D_{\eta } \big) \Big] ^{n} \xi , 
\hspace{5mm} 
\widetilde{F} \widetilde{\xi } \equiv \sum_{n=0}^{\infty } \Big[ \tilde{\xi } \big( \teta - D_{\teta } \big) \Big] ^{n} \tilde{\xi } , 
\end{align}
which satisfy $D_{\eta } \, F \xi + F \xi \, D_{\eta } = 1$ and $D_{\teta } \, \widetilde{F} \tilde{\xi } + \widetilde{F} \tilde{\xi } \, D_{\teta } = 1$, respectively.\footnote{If you prefer, you can use the coalgebraic notation: $F \xi ( A ) = \pi_{1} \, \widehat{\bf G} [ e^{{\bf G}^{-1}(\Psi _{\eta \teta }) } \wedge \pi_{1} \, \xi \, \widehat{\bf G}^{-1} (e^{\wedge \Psi _{\eta \teta } } \wedge A )]$\,.
The author thanks to T.Erler for comments.}  
See also \cite{Goto:2015pqv, Kunitomo:2015usa, GK, Matsunaga:2015kra}. 
These $F \xi $ and $\widetilde{F} \tilde{\xi }$ consist of the pure-gauge-like (functional) field $\Psi _{\eta \teta } [ \varphi ]$ and operators ${\bf L}^{\boldsymbol{\eta }}$, ${\bf L}^{\boldsymbol{\teta }}$, $\eta$, $\teta $, $\xi $ and $\tilde{\xi }$. 
Using these pieces, one can construct $\Psi _{\mathbf{D}} [\varphi ]$ via $\Psi _{\eta {\rm D}} [\varphi ]$ and $\Psi _{{\rm D} \teta }[\varphi ]$ as follows, 
\begin{align*}
\Psi _{\mathbf{D}} [\varphi ] \equiv F \xi \, \Psi _{{\rm D} \teta} [\varphi ] = - \widetilde{F} \tilde{\xi } \, \Psi _{\eta {\rm D}} [\varphi ] . 
\end{align*}
This $\Psi _{\mathbf{D}}$ quickly satisfies (\ref{WZW 2}), and thus, for example, one can check that (\ref{pure eq}) holds up to $D_{\eta }$-exacts and $D_{\teta }$-exacts in large-space parametrisation. 
Note that as well as that of the NS sector, the form of $F$ or $\widetilde{F}$ is not unique. 
In the NS-NS sector, this type of ambiguities of (\ref{F}) can be crossed over between left-moving and right-moving sectors. 
Although $F \xi $ and $\widetilde{F} \tilde{\xi }$ do not exactly commute under the above choice of (\ref{F}) and the equality holds up to $D_{\eta }$-exacts or $D_{\teta }$-exacts, we can have the strict commutativity and equality, which we see in the next section.

%\clearpage 

\section{Properties}

\subsection*{Single functional form} 

As we found, two or more types of functional fields $\Psi _{\eta \teta } [\varphi ]$, $\Psi _{\mathbf{D}} [\varphi ]$ appear in the WZW-like action (\ref{Action}). 
Their algebraic relations make computations easy, but, at the same time, give constraints on these functional fields: 
the existence of many types of (functional) fields satisfying constraint equations would complicate its gauge fixing problem. 
It is known that in the NS sector, (alternative) WZW-like actions have single functional forms \cite{Matsunaga:2015kra}. 
We show as well as NS actions, our NS-NS action $S_{\eta \teta } [\varphi ]$ has a single functional form which consists of the single functional $\Psi _{\eta \teta } [\varphi ]$ and elementally operators. 
It may be helpful in the gauge fixing problem. 

\vspace{2mm} 

Recall that in the left-and-right large Hilbert space $\cH $ of the NS-NS sector, because of $\eta \, \xi + \xi \, \eta = 1$ and $\teta \, \tilde{\xi } + \tilde{\xi } \, \teta = 1$, the $\eta $-complex and $\teta $-complex are both exact: 
\begin{align*}
\dots \overset{\eta }{\longrightarrow } \cH \overset{\eta }{\longrightarrow } \cH \overset{\eta }{\longrightarrow } \cH \overset{\eta }{\longrightarrow } \dots \hspace{2mm} ({\rm exact}) \, , \hspace{5mm} 
%\no 
\dots \overset{\teta }{\longrightarrow } \cH \overset{\teta }{\longrightarrow } \cH \overset{\teta }{\longrightarrow } \cH \overset{\teta }{\longrightarrow } \dots  \hspace{2mm} ({\rm exact}) . 
\end{align*} 
Furthermore, since $\eta \, \teta \, + \teta \, \eta = 0$, $\eta \, \tilde{\xi } + \tilde{\xi } \, \eta = 0$, $\teta \, \xi + \xi \, \teta = 0$, and $\xi \, \tilde{\xi } + \tilde{\xi } \, \xi = 0$ hold, we have the direct sum decomposition of the large state space $\cH $ as follows: 
\begin{align*}
\cH = \eta \, \teta \, \cH \oplus \eta \, \tilde{\xi } \, \cH \oplus \teta \, \xi \, \cH \oplus \xi \tilde{\xi } \, \cH .  
\end{align*}

Likewise, the existence of (\ref{F}) satisfying $D_{\eta } \, F \xi + F \xi \, D_{\eta } = 1$ and $D_{\teta } \, \widetilde{F} \tilde{\xi } + \widetilde{F} \tilde{\xi } \, D_{\teta } = 1$ implies that the both $D_{\eta }$-complex and $D_{\teta }$-complex are also exact in this large state space $\cH $: 
\begin{align*}
\dots \overset{D_{\eta }}{\longrightarrow } \cH \overset{D_{\eta }}{\longrightarrow } \cH \overset{D_{\eta }}{\longrightarrow } \cH \overset{D_{\eta }}{\longrightarrow } \dots \hspace{2mm} ({\rm exact}) \, , \hspace{5mm} 
%\no 
\dots \overset{D_{\teta }}{\longrightarrow } \cH \overset{D_{\teta }}{\longrightarrow } \cH \overset{D_{\teta }}{\longrightarrow } \cH \overset{D_{\teta }}{\longrightarrow } \dots  \hspace{2mm} ({\rm exact}) . 
\end{align*} 
However, we saw (\ref{F}) do not exactly commute each other. 
Does there exist a direct sum decomposition using these exact sequences? 
To achieve this, we consider  
\begin{align*}
\cF \equiv \sum_{n=0 }^{\infty } \big[ \widetilde{F} \xi ( \eta \widetilde{F}^{-1} - \widetilde{F}^{-1} D_{\eta } ) \big] ^{n} \widetilde{F} , 
\hspace{5mm} 
\cF ^{-1} \equiv \eta \xi \widetilde{F}^{-1} + \xi \widetilde{F}^{-1} D_{\eta } \, . 
\end{align*}
One can quickly find that as well as (\ref{F}), this $\cF $ and its inverse $\cF ^{-1}$ also provide 
\begin{align*}
D_{\eta } = \cF \, \eta \, \cF ^{-1} , \hspace{5mm} 
D_{\teta } = \cF \, \teta \, \cF ^{-1}  , 
\end{align*}
and it makes possible to have the following decompositions of the identity, 
\begin{align*}
D_{\eta } \, \cF _{\xi } + \cF _{\xi } \, D_{\eta } = 1 \, , \hspace{3mm} 
D_{\teta } \, \cF _{\tilde{\xi }} + \cF _{\tilde{\xi }} \, D_{\teta } = 1 \,,
\hspace{3mm} 
( \, \cF _{\xi } \equiv \cF \, \xi \, \cF ^{-1} , \hspace{2mm} 
\cF _{\tilde{\xi }} \equiv \cF \, \widetilde{\xi } \, \cF ^{-1} \, ) \, . 
\end{align*}
Furthermore, now, these operators all are constructed from single $\cF $, we have 
\begin{align*}
D_{\eta } \, D_{\teta } + D_{\teta } \, D_{\eta }= 0, \hspace{5mm} 
D_{\eta } \, \cF _{\tilde{\xi }} + \cF _{\tilde{\xi }} \, D_{\eta } = 0 , \hspace{5mm}   
D_{\teta } \, \cF _{\xi } + \cF _{\xi } \, D_{\teta } = 0 , \hspace{5mm}   
\cF _{\xi } \, \cF _{\tilde{\xi }} + \cF _{\tilde{\xi }} \, \cF _{\xi } = 0 , 
\end{align*}
which give us the desired direct sum decomposition of the large state space $\cH $\,: 
\begin{align*} 
\cH = D_{\eta } \, D_{\teta } \, \cH \oplus D_{\eta } \, \cF _{\tilde{\xi }} \, \cH \oplus D_{\teta } \, \cF _{\xi } \, \cH \oplus \cF _{\xi } \, \cF _{\tilde{\xi }} \, \cH \, . 
\end{align*}

Since $Q \Psi _{\eta \teta } = D_{\eta } F_{\xi } D_{\teta } F_{\tilde{\xi }} ( Q \Psi _{\eta \teta } )$ and $D_{\teta } D_{\eta } \Psi _{t} = \partial _{t} \Psi _{\eta \teta }$, using this $\cF $, we find
\begin{align}
\label{F form}
S_{\eta \teta } [\varphi ] & = \int_{0}^{1} dt \, \big{\langle } \Psi _{t} [ \varphi (t) ] , \, Q \, \Psi _{\eta \teta } [\varphi (t) ] \big{\rangle } 
\no 
& = \int_{0}^{1} dt \, \big{\langle }  \partial _{t} \Psi _{\eta \teta } [\varphi (t) ] , \, \cF _{\xi } \cF _{\tilde{\xi } } \, Q \, \Psi _{\eta \teta } [ \varphi (t) ] \big{\rangle } . 
\end{align}
It consists of the single functional $\Psi _{\eta \teta } [\varphi ]$ and elementary operators ${\bf L}^{\boldsymbol{\eta }}$, ${\bf L}^{\boldsymbol{\teta }}$, $\eta $, $\xi $, $\teta $, $\tilde{\xi }$, and $Q$. 
One can also check that this (\ref{F form}) has topological $t$-dependence using (\ref{WZW 2}) and the commutation relation $\ld \mathbf{D} , F \xi \rd = - F \xi \ld \mathbf{D} , D_{\eta } \rd F \xi + \ld D_{\eta } , F \xi \, \mathbf{D} \, F \xi \rd $\,. 

\subsection*{Equivalence of two constructions} 

In section 4, we presented two constructions of the WZW-like action. 
We explain these two actions are equivalent and derive a field redefinition connecting these. 
By construction, the equivalence of $S_{\eta \teta } [ \Phi ]$ and $S_{\eta \teta } [\Psi ]$ follows if we consider the identification 
\begin{align}
\Psi _{\eta \teta } [\Phi ] \cong \Psi _{\eta \teta } [ \Psi ] \, . \label{identify} 
\end{align}
It is trivial from the fact that the WZW-like action (\ref{Action}) has the single functional form (\ref{F form}) which consists of $\Psi _{\eta \teta }$ and elementally operators. 
Since both actions have the same WZW-like structure, one can impose this identification and solve it as a field relation. 
See also \cite{Goto:2015pqv, Matsunaga:2015kra, Erler:2015rra, Erler:2015uba, Erler:2015uoa}. 

\niu{Field relation} 

Note that the identification of states (\ref{identify}) provides the identification of their Fock spaces 
\begin{align*}
e^{\wedge \Psi _{\eta \teta } [\Phi ] } = e^{\wedge \Psi _{\eta \teta } [\Psi ] } \, ,
\end{align*}
Under the identification (\ref{identify}), by acting $\partial _{t}$, we have 
\begin{align}
\label{identify sub}
\Psi _{t } [ \Phi ] = \Psi _{t} [ \Psi ]  + D_{\eta } \mathchar`-  {\rm exacts} + D_{\teta } \mathchar`-  {\rm exacts} . 
\end{align}
Note that these $D_{\eta }$-exact or $D_{\teta }$-exact term does not contribute in the action. 
We thus consider 
\begin{align*}
e^{\wedge \Psi _{\eta \teta } [ \Psi (t) ] } \wedge \Psi _{t} [\Psi  (t)] & = e^{\wedge \Psi _{\eta \teta } [\Phi (t) ] } \wedge \Psi _{t} [\Phi (t) ] 
=  \widehat{\bf G} \Big( e^{\wedge \Phi (t) } \wedge \xi \tilde{\xi } \partial _{t} \Phi (t) \Big) .
\end{align*}
The ambiguity appearing in (\ref{identify sub}) is completely absorbed into the gauge transformations: 
\begin{align*}
\delta \Big( e^{\wedge \Psi _{\eta \teta } [\varphi ]} \Big) = e^{\wedge \Psi _{\eta \teta } [\varphi ] } \wedge \xi \tilde{\xi } \delta \Psi _{\eta \teta } [\varphi ] = e^{\wedge \Psi _{\eta \teta } [\varphi ] } \wedge \xi \tilde{\xi } \Big( Q \Lambda + D_{\eta } \Omega + D_{\teta } \widetilde{\Omega } \Big) .
\end{align*}
Since cohomomorphism $\widehat{\bf G}$ is invertible, we obtain the following field relation 
\begin{align*}
\Phi & = - \pi _{1} \boldsymbol{\eta } \, \boldsymbol{\teta } \, \int_{0}^{1} dt \, \widehat{\bf G}^{-1} \Big( e^{\wedge \Psi _{\eta \teta } [ \Psi (t) ] } \wedge \Psi _{t} [\Psi (t) ] \Big)
\no & 
= \pi _{1} \int_{0}^{1} dt \, \widehat{\bf G}^{-1} \Big( e^{\wedge \Psi _{\eta \teta } [ \Psi (t) ] } \wedge D_{\teta } D_{\eta } \Psi _{t} [\Psi (t) ] \Big) \, . 
\end{align*}
By using the WZW-like relation (\ref{WZW 2}), it reduces to the following expression 
\begin{align*} 
\Phi = \pi _{1} \int_{0}^{1} dt \, \widehat{\bf G}^{-1} \Big( \boldsymbol{\partial _{t} } \, e^{\wedge \Psi _{\eta \teta } [ \Psi (t) ] } \Big)
%\no & 
= \pi _{1} \widehat{\bf G}^{-1} \Big( e^{\wedge \Psi _{\eta \teta } [ \Psi ] } \Big) \, , 
\end{align*}
which can be directly derived from (\ref{identify}). 

\subsection*{Relation to $L_{\infty }$ theory} 

We write $\Phi$ for the small-space dynamical string field considered in section 4, and write $\Phi '$ for the dynamical string field of the $L_{\infty }$ action proposed in \cite{Erler:2014eba}. 
As well as $\Phi$, this $\Phi '$ belongs to the small Hilbert space: $\eta \, \Phi ' = 0$ and $\teta \, \Phi ' = 0$. 
Recall that using the small-space dynamical string field $\Phi $, we constructed an action 
\begin{align*}
S_{\eta \teta } [\Phi ] & = \int_{0}^{1} dt \, \big{\langle } \pi _{1}  \widehat{\bf G} \Big( \xi \tilde{\xi } \partial _{t} \Phi (t) \wedge e^{\wedge \Phi (t)} \Big)  , \, Q \, \pi _{1}  \widehat{\bf G} \Big( e^{\wedge \Phi (t) } \Big) \big{\rangle } \, . 
\end{align*}
We will show that this $S_{\eta \eta } [\Phi ]$ is exactly off-shell equivalent to the $L_{\infty }$ action, 
\begin{align}
S_{L_{\infty }} [ \Phi ' ] = \frac{1}{2} \big{\langle } \xi \tilde{\xi } \Phi ' , \, Q \Phi ' \big{\rangle }  + \sum_{n=1}^{\infty } \frac{1}{(n+1)!} \langle \xi \tilde{\xi } \Phi ' , L_{n+1} ( \overbrace{\Phi ' , \dots , \Phi ' }^{n} , \Phi ' ) \big{\rangle } \, .  \label{EKS}
\end{align}
Let $\Phi '(t)$ be a path connecting $\Phi '(0) = 0$ and $\Phi '(1) = \Phi '$, where $t \in [0,1]$ is a real parameter. 
We write $S_{L_{\infty }} [ \Phi '(t) ]$ for the function given by replacing $\Phi '$ of (\ref{EKS}) with $\Phi '(t)$, which satisfies $S_{L_{\infty }} [ \Phi '(1) ] = S_{L_{\infty }} [\Phi ']$ and $S_{L_{\infty }}[ \Phi '(0) ] = S_{L_{\infty }} [0] = 0$. 
Then, we have 
\begin{align*} 
S_{L_{\infty }} [\Phi ' ] = \int_{0}^{1} dt \, \frac{d}{d t} \, S_{L_{\infty } } [\Phi '(t) ] = \int_{0}^{1} dt \, \big{\langle } \xi \tilde{\xi } \partial _{t} \Phi '(t) , \, \pi _{1} {\bf L}^{\rm NS,NS} \, e^{\wedge \Phi '(t) } \big{\rangle } \, . 
\end{align*}
Using coalgebraic notation and ${\bf L}^{\rm NS,NS} = \widehat{\bf G}^{-1} \, {\bf Q} \,  \widehat{\bf G}$, we find  
\begin{align*} 
S_{L_{\infty }} [\Phi ' ]  
& = \int_{0}^{1} dt \, \big{\langle } \pi _{1} \Big( \xi \tilde{\xi } \partial _{t} \Phi '(t) \wedge e^{\wedge \Phi '(t) } \Big) , \, \pi _{1}  \widehat{\bf G}^{-1} \, {\bf Q} \,  \widehat{\bf G} \Big( e^{\wedge \Phi '(t) } \Big) \big{\rangle } 
\no 
& = \int_{0}^{1} dt \, \big{\langle } \pi _{1}  \widehat{\bf G} \Big( \xi \tilde{\xi } \partial _{t} \Phi '(t) \wedge e^{\wedge \Phi '(t)} \Big)  , \, Q \, \pi _{1}  \widehat{\bf G} \Big( e^{\wedge \Phi '(t) } \Big) \big{\rangle } \, . 
\end{align*}
In the second equality, we used the fact that $\widehat{\bf G}$ is a cyclic $L_{\infty }$-isomorphism compatible with the BPZ inner product. 
This just gives one realization of our WZW-like action (\ref{Action}) in small-space parametrisation. 
Hence, with the (trivial) identification of the string fields, 
\begin{align*}
\Phi \cong \Phi ' \, , 
\end{align*}
we obtained a proof that the $L_{\infty }$ action $S_{L_{\infty }} [\Phi ']$ proposed in \cite{Erler:2014eba} is equivalent to our $S_{\eta \teta } [\Phi ]$. 
It implies that since $S_{\eta \teta }[\Psi ]$ has the same WZW-like structure as $S_{\eta \teta }[\Phi ]$, WZW-like actions $S_{\eta \teta } [\Phi ]$ and $S_{\eta \teta }[\Psi ]$ both are equivalent to that of $L_{\infty }$ formulation. 
See also \cite{Goto:2015pqv, Erler:2015uoa}

\niu{WZW-like reconstruction of $L_{\infty }$ action} 

In the $L_{\infty }$ action, the $L_{\infty }$ triplet is given by $( \boldsymbol{\eta } , \boldsymbol{\teta } \, ; {\bf L}^{\rm NS,NS} )$. 
We thus consider a functional $\Phi _{\eta \teta } [ \varphi ]$ which satisfies two constraint equations defined by $\boldsymbol{\eta }$ and $\boldsymbol{\teta }$, 
\begin{subequations} 
\begin{align}
 \pi _{1} \, \boldsymbol{\eta } \, ( e^{\wedge \Phi _{\eta \teta } [\varphi ] } ) = \eta \, \Phi _{\eta \teta } [\varphi ] = 0 ,
\\ 
\pi _{1} \, \boldsymbol{\teta } \, ( e^{\wedge \Phi _{\eta \teta } [\varphi ] } ) = \teta \, \Phi _{\eta \teta } [\varphi ] = 0 . 
\end{align}
\end{subequations} 
By acting derivation $\mathbf{D}$ satisfying both $\ld \mathbf{D} , \eta \rd = 0$ and $\ld \mathbf{D} , \teta \rd = 0$ on these, we find $\eta \, ( \mathbf{D} \Phi _{\eta \teta } ) = 0$ and $\teta \, ( \mathbf{D} \Phi _{\eta \teta } ) = 0$. 
It implies that with some functional $\Phi _{\mathbf{D}} [\varphi ]$, we have the WZW-like relation, 
\begin{align}
(-)^{\mathbf{D}} \mathbf{D} \, \Phi _{\eta \teta } [ \varphi ] = - \eta \, \teta \, \Phi _{\mathbf{D}} [\varphi ] . 
\end{align}
The existence of $\Phi _{\mathbf{D}}$ is ensured because $\eta$-complex and $\teta$-complex are both exact in the left-and-right large Hilbert space. 
Using $\Phi _{\eta \teta } [\varphi ]$, we can consider the Maurer-Cartan element for the remaining $L_{\infty }$ products ${\bf L}^{\rm NS,NS}$\,: 
\begin{align*}
\pi _{1} \, {\bf L}^{\rm NS,NS} ( e^{\Phi _{\eta \teta } [\varphi ]} ) 
= Q \, \Phi _{\eta \teta } [\varphi ] + \sum_{n=2}^{\infty } \frac{1}{n!} L_{n} \big( \overbrace{ \Phi _{\eta \teta } [\varphi ] \, , \dots , \Phi _{\eta \teta } [\varphi ] }^{n} \big) \, . 
\end{align*}
Note that there also exists an associated field $\Phi _{L} [\varphi ]$ such that 
\begin{align*}
\pi _{1} \, {\bf L}^{\rm NS,NS} ( e^{\wedge \Phi _{\eta \teta }[ \varphi ]} ) = \eta \, \teta \, \Phi _{L} [\varphi ] \, .
\end{align*}
According to our recipe, utilizing these ingredients, we can construct a WZW-like action\footnote{The NS-NS actions given by \cite{Sen:2015uaa , Jurco:2013qra} also has this kind of WZW-like structure and WZW-like form of the action. Its $L_{\infty }$ triplet is quickly obtained by replacing ${\bf L}^{\rm NS,NS}$ of $(\boldsymbol{\eta } , \boldsymbol{\teta} \,; {\bf L}^{\rm NS,NS} )$ with the $L_{\infty }$ products appearing the action of \cite{Sen:2015uaa, Jurco:2013qra} because of their small-space constraints.}: 
\begin{align}
\label{WZW L action}
S_{L_{\infty }} [\varphi ] & = \int_{0}^{1} dt \, \big{\langle } \Phi _{t} [ \varphi (t) ] , \, \pi _{1} \, {\bf L}^{\rm NS,NS} ( e^{\wedge \Phi _{\eta \teta } [ \varphi (t) ] } ) \big{\rangle } 
\no 
& = \int_{0}^{1} dt \, \big{\langle } \Phi _{t} [ \varphi (t) ] , \, \eta \, \teta \, \Phi _{L} [ \varphi (t) ] \big{\rangle } \, .
\end{align}
One can check this action (\ref{WZW L action}) has topological $t$-dependence and gauge invariance in the WZW-like manner. 
In particular, since $\eta$ and $\teta$ are linear $L_{\infty }$ products, their shifted products are themselves. 
Thus, one can compute it with truncated versions of (\ref{1st term}) or (\ref{2nd term}). 
We notice that if we set $\varphi = \Phi$ satisfying $\eta \, \Phi = \teta \, \Phi = 0$, it naturally induces a trivial form of the functional, $\Phi _{\eta \teta } [\Phi ] \equiv \Phi $, because of the triviality of $\eta $- and $\teta $-cohomology. 
Similarly, if we use $\varphi = \Psi$, it also implies $\Phi _{\eta \eta } [\Psi ] \equiv \eta \teta \, \Psi$\,. 
While its small-space parametrisation is just the $L_{\infty }$ action given by \cite{Erler:2014eba}, its large-space parametrisation is just a trivial up-lift of small-space one.

\vspace{2mm} 

\niu{Off-shell duality of $L_{\infty }$ triplets} 

As we mentioned, when $\widehat{\bf G}$ is cyclic in the BPZ inner product, (\ref{Duality}) ensures not only the equivalence of $L_{\infty }$ triplets but also the off-shell equivalence of resultant WZW-like actions.   
To see this, it is useful to consider the Maurer-Cartan-like element in {\it the correlation function}\,: 
\begin{align*}
\big{\langle } \mathcal{MC}_{\alpha } ( \cA ) \big{\rangle } \equiv \sum_{n=1}^{\infty } \frac{1}{(n+1)!} \big{\langle } \cA \, , \big[ \overbrace{\cA , \dots , \cA}^{n} \big] ^{\alpha } \big{\rangle } \, . 
\end{align*}
Note that the above sum starts from $n=1$, namely, two-inputs is the lowest. 
In the correlation function $\langle \, \dots \rangle$, the BPZ cyclic property of $\widehat{\bf G}$ is just $\langle \widehat{\bf G} ( \, \dots  ) \rangle = \langle \, \dots  \rangle $. 
We thus obtain 
\begin{align}
\label{Off-shell duality}
\big{\langle } \mathcal{MC}_{Q} ( \cA ) \big{\rangle } = \big{\langle } \widehat{\bf G}^{-1} \cdot \mathcal{MC}_{Q } ( \cA ) \big{\rangle } = \big{\langle } \mathcal{MC}_{L } ( \cA ' ) \big{\rangle } \, ,
\end{align} 
where $\mathcal{MC}_{L} (\cA ')$ is the Maurer-Cartan element for ${\bf L}^{\rm NS,NS}$ and $\cA '$ is a state satisfying dual constraints for $\cA$\,. 
Note that when the state $\cA$ satisfies $\mathcal{MC}_{L^{\eta}} (\cA) = \mathcal{MC}_{L^{\teta }}(\cA )= 0$, the state $\cA '$ satisfies $\mathcal{MC}_{\eta } (\cA ' ) = \mathcal{MC}_{\teta } (\cA ' ) = 0$. 

\vspace{2mm}

Let us introduce a Grassmann variable $\tilde{t}$ satisfying $(\, \tilde{t}\, )^{2} =0$, and write $\cA [\varphi ] \equiv \Psi _{\eta \teta }[\varphi ] + \tilde{t} \, \Psi _{t} [\varphi ]$. 
Using a measure factor $d \equiv dt \cdot \partial _{\,\tilde{t}}$\,, we can express the WZW-like action (\ref{Action}) as 
\begin{align} 
S_{\eta \teta } = \int d \, \big{\langle } \mathcal{MC}_{Q} ( \cA ) \big{\rangle } \, , 
\end{align} 
which reminds us the Chern-Simons form and its geometrical quantity. 
Likewise, using $\cA ' [\varphi ] \equiv \Phi _{\eta \teta } [\varphi ] + \tilde{t} \, \Phi _{t} [\varphi ]$, the WZW-likely extended $L_{\infty }$ action (\ref{WZW L action}) can be written as 
\begin{align} 
S_{L_{\infty }} = \int d \, \big{\langle } \mathcal{MC}_{L} ( \cA ' ) \big{\rangle } \, . 
\end{align} 
Then, the equality (\ref{Off-shell duality}) of the Maure-Cartan elements in the correlation function concludes the off-shell equivalence between our WZW-like action (\ref{Action}) based on the $L_{\infty }$ triplet $({\bf L}^{\boldsymbol{\eta }} , {\bf L}^{\boldsymbol{\teta }} \,; {\bf Q})$ and the (WZW-likely extended) $L_{\infty }$ action (\ref{WZW L action}) based on the $L_{\infty }$ triplet $(\boldsymbol{\eta } , \boldsymbol{\teta } \,; {\bf L}^{\rm NS,NS})$\,. 
Note that this off-shell equivalence does not necessitate detailed information about dynamical string fields. 
It is a powerful and significant consequence of the WZW-like structure. 

\subsection*{Relation to the earlier WZW-like theory} 

The $L_{\infty }$ triplet of the earlier WZW-like action is given by $( {\bf L}^{-, \rm NS} ,\boldsymbol{\teta } \, ; \boldsymbol{\eta } )$. 
In this WZW-like NS-NS theory of \cite{Matsunaga:2014wpa}, a solution of both Maurer-Cartan equations for ${\bf L}^{-,\rm NS}$ and ${\boldsymbol{\teta }}$ plays the most important role. 
We write $\varphi '$ for a dynamical NS-NS string field and consider a functional $\cG _{L} = \cG _{L} [\varphi ']$ of this string field. 
Let $\cG _{L}$ be a state which has ghost number $2$, left-moving picture number $0$, and right-moving picture number $-1$ state in the large Hilbert space. 
When this $\cG _{L}$ satisfies 
\begin{subequations}
\begin{align} 
Q \, \cG _{L} + \sum_{n=1}^{\infty } \frac{1}{(n+1)!} \big[ \overbrace{\cG _{L} \, , \dots , \cG _{L} }^{n} \, , \cG _{L} \big] ^{-, \rm NS}  & = 0 , \label{c1}
\\ 
\teta \, \cG _{L} & = 0 ,  \label{c2}
\end{align}
\end{subequations} 
we call $\cG _{L}$ {\it a pure-gauge-like (functional) field}. 
Let $\mathbf{D}$ be a derivation operator of ${\bf L}^{-,\rm NS}$ and ${\boldsymbol{\teta }}$\,: 
Namely $\mathbf{D} \, {\bf L}^{-,\rm NS} - (-)^{\mathbf{D}} {\bf L}^{-,\rm NS} \, \mathbf{D} = 0$ and $\mathbf{D} \, \boldsymbol{\teta } - (-)^{\mathbf{D}} \boldsymbol{\teta } \, \mathbf{D} = 0$. 
For example, one can take $\mathbf{D} = \eta $, $\partial _{t}$, and $\delta $. 
Once the above pure-gauge-like (functional) field $\cG _{L}$ is given, we consider 
\begin{align}
(-)^{\mathbf{D}} \mathbf{D} \, \cG _{L} = - Q_{\cG _{L}} \teta \, \Psi '_{\mathbf{D}} , \label{cWZW}
\end{align} 
which we call {\it the (earlier) WZW-like relation}. 
Here, $\Psi '_{\mathbf{D}} = \Psi '_{\mathbf{D}} [\varphi ']$ is a functional of the dynamical string field, which has the same ghost, left-moving-picture, and right-moving-picture numbers as $d$. 
We call this $\Psi '_{\mathbf{D}}[\varphi ']$ satisfying (\ref{cWZW}) as {\it an associated (functional) field}. 
Note that $Q_{\cG _{L}}$, the first $\cG _{L}$-shifted ${\bf L}^{-,\rm NS}$, satisfies $Q_{\cG _{L}} \teta + \teta \, Q_{\cG _{L} } = 0$ because of (\ref{c1}) and (\ref{c2}). 

In \cite{Matsunaga:2014wpa}, using these $\cG _{L} [\varphi ']$ and $\Psi _{\mathbf{D}} [\varphi ]$, a WZW-like action was given by 
\begin{align}
\label{cWZW Action}
S [\varphi ] = \int_{0}^{1} dt \, \big{\langle } \Psi '_{t} [\varphi '(t) ] , \, \eta \, \cG _{L} [ \varphi '(t) ] \big{\rangle } . 
\end{align}
We write $\Psi '_{t}[ \varphi '(t) ]$ for the associated field $\Psi '_{\mathbf{D}}[\varphi '(t) ]$ with $\mathbf{D} = \partial _{t}$, and $\varphi '(t)$ is a path connecting $\varphi '( 0 ) = 0$ and $\varphi '(1) = \varphi '$, where $t \in [0 , 1]$ is a real parameter. 
While the dynamical string field is taken $\varphi ' = \Psi '$ in the left-and-right large Hilbert space in \cite{Matsunaga:2014wpa}, if one prefer, one can consider the small-space parametrisation. 
But now, we would like to focus on its WZW-like structure. 

\vspace{2mm}

By its construction, we notice that the situation is parallel to the NS sector of heterotic string field theory \cite{Berkovits:2004xh}: 
Unfortunately, as \cite{Goto:2015hpa}, we do not have exact off-shell equivalence at all order but only have lower order equivalence. 
For example, by taking the following nonlinear partially gauge-fixing condition on $\varphi ' = \Psi ' $ with the small-space string field $\Phi $, 
\begin{align*}
\Psi ' &= \tilde{\xi } \Big{\{ } \xi \Phi + \frac{1}{3!} \xi L_{2}^{\rm -,NS}\big( \xi \Phi , \Phi \big)  
+ \frac{1}{4!} \Big( \xi L_{3}^{\rm -,NS} \big( Q \xi \Phi , \xi \Phi , \Phi \big) + \xi L_{3}^{\rm - , NS} \big( X \Phi , \xi \Phi , \Phi \big) \Big) 
\no & \hspace{15mm} 
+ \frac{1}{4!} \Big( \frac{4}{3} \xi L_{2}^{\rm -,NS} \big( \Phi , \xi L_{2}^{\rm -,NS} ( \xi \Phi , \Phi )  \big) + \frac{1}{3} \xi L_{2}^{\rm -,NS} \big( \xi \Phi , \xi L_{2}^{\rm -,NS} ( \Phi , \Phi ) \big) 
\no & \hspace{30mm}
- \frac{2}{3} \xi L_{2}^{\rm -,NS} \big( \xi \Phi , L_{2}^{\rm -,NS} ( \xi \Phi , \Phi ) \big) \Big) \Big{\} } 
+ \dots \, , 
\end{align*} 
the action (\ref{cWZW Action}) reduces to the $L_{\infty }$ action based on their asymmetric construction of \cite{Erler:2014eba}. 
Hence, WZW-like actions (\ref{Action}) and (\ref{cWZW}) relate each other via field redefinitions, at least lower order. 

%\clearpage 

\section{Conclusion}

We presented that a triplet of mutually commutative $L_{\infty }$ products $({\bf L}^{c} , {\bf L}^{\tilde{c}} \, ; {\bf L}^{p})$ completely determine the gauge structure of the WZW-like action. 
As we showed, every known NS-NS superstring field theory \cite{Goto:2015pqv, Erler:2014eba, Matsunaga:2014wpa, Jurco:2013qra, Sen:2015uaa} potentially have the following WZW-like structure and WZW-like form of the action, which is one interesting result: 
By using two of it as constraint equations, 
\begin{subequations}
\begin{align}
\pi _{1} \, {\bf L}^{c} \, e^{\wedge \Psi _{c \tilde{c}} [\varphi ]} = \sum_{n=1} \frac{1}{n!} L_{n}^{c} \big( \Psi _{c \tilde{c}} [\varphi ] , \dots , \Psi _{c \tilde{c} }[\varphi ] \big) = 0 ,  
\\ 
\pi _{1} \, {\bf L}^{\tilde{c}} \, e^{\wedge \Psi _{c \tilde{c}} [\varphi ]} = \sum_{n=1} \frac{1}{n!} L_{n}^{\tilde{c}} \big( \Psi _{c \tilde{c}} [\varphi ] , \dots , \Psi _{c \tilde{c} }[\varphi ] \big) = 0 ,  
\end{align}
\end{subequations}
and introducing a functional $\Psi _{c \tilde{c}} [\varphi ]$ of some dynamical string field $\varphi $ satisfying these constraints, we constructed a gauge-invariant WZW-like action for the NS-NS superstring field theory, 
\begin{align}
S_{c \tilde{c}} [\varphi ] = \int d \, \big{\langle }  \mathcal{MC}_{L^{p}} ( \cA ) \big{\rangle } = \int _{0}^{1} dt \, \big{\langle } \Psi _{t} [\varphi (t)] , \pi _{1} {\bf L}^{p} \, e^{\wedge \Psi _{c \tilde{c}} [\varphi (t)]} \big{\rangle } \, ,
\end{align}
whose on-shell condition is given by the Maurer-Cartan element of the other $L_{\infty }$, 
\begin{align} 
\pi _{1} \, {\bf L}^{p} \, e^{\wedge \Psi _{c \tilde{c}} [\varphi ]} = \sum_{n=1} \frac{1}{n!} L_{n}^{p} \big( \Psi _{c \tilde{c}} [\varphi ] , \dots , \Psi _{c \tilde{c} }[\varphi ] \big) = 0 \, . 
\end{align} 
One can prove its gauge invariance using the functional $\Psi _{c \tilde{c}} [\varphi ]$ and algebraic relations derived from the mutual commutativity of the $L_{\infty }$ triplet $({\bf L}^{c} , {\bf L}^{\tilde{c}} \, ; {\bf L}^{p})$,\footnote{As we found, in the NS-NS sector, one or two $L_{\infty }$ of the triplet becomes linear. However, in general, all $L_{\infty }$ of the triplet can be nonlinear: When we include the Ramond sectors, it will be the case, which is expected from the result of \cite{Matsunaga:2015kra}. Actually, with deep insights, one can find a pair of (nonlinear) $A_{\infty }$ products plays such a role in WZW-like actions for open superstring field theory including the NS and R sectors \cite{Erler}. } without using details of the dynamical string field $\varphi $. 
Since each know NS-NS action has its WZW-like form, one can say that to study its $L_{\infty }$ triplet is equivalent to know the gauge structure of NS-NS superstring field theory. 
In this paper, we focused on two $L_{\infty }$ triplets $({\bf L}^{\boldsymbol{\eta }} , {\bf L}^{\boldsymbol{\teta}} \,;{\bf Q})$ and $(\boldsymbol{\eta } , \boldsymbol{\teta} \,; {\bf L}^{\rm NS,NS})$\, which provide the $L_{\infty }$ action of \cite{Erler:2014eba}. 
Particularly, we presented detailed analysis of the former and proved their off-shell equivalence with several general or exact results. 
We also discussed the relation to the earlier WZW-like action of \cite{Matsunaga:2014wpa}. 
We showed as well as the WZW-like action of the NS sector, our WZW-like action of the NS-NS sector has a single functional form, which may be a new approach to the gauge-fixing problem of WZW-like theory. 

\section*{Acknowledgments} 

The author would like to thank Theodore Erler, Keiyu Goto, Hiroshi Kunitomo, and Martin Schnabl. 
The author also thank the referee of the previous work, JHEP {\bf 1701} (2017) 022, for reminding him about this topic. 
This research has been supported by the Grant Agency of the Czech Republic, under the grant P201/12/G028. 

\appendix 

\section{General WZW-like action based on $(\bL ^{c} , \bL ^{\tilde{c}} \,; \bL ^{p})$} 

In section 6, we gave the general WZW-like action based on a general $L_{\infty }$ triplet $(\bL ^{c} , \bL ^{\tilde{c}} \,; \bL ^{p})$\,. 
In this appendix, we prove that the general WZW-like action, 
\begin{align*}
S_{c\tilde{c}} [\varphi ] = \int _{0}^{1} dt \, \la \Psi _{t} [\varphi (t)] , \, \pi _{1} {\bf L}^{p} e^{\wedge \Psi _{c\tilde{c} } [\varphi  (t)] } \ra \, , 
\end{align*}
has topological parameter dependence: 
Its variation is given by 
\begin{align*}
\delta S_{c\tilde{c}} [\varphi ] = \la \Psi _{\delta } [\varphi ] , \, \pi _{1} {\bf L}^{p} e^{\wedge \Psi _{c\tilde{c} } [\varphi ] } \ra \, . 
\end{align*}
Then, because of the nilpotency of $L_{\infty }$ triplet $(\bL ^{c} , \bL ^{\tilde{c}} \,; \bL ^{p})$\,, the general WZW-like action is invariant under the gauge transformations generated by $\bL ^{c}$\,, $\bL ^{\tilde{c}}$, and $\bL ^{p}$\,, 
\begin{align*}
\Psi _{\delta } [\varphi ] = \pi _{1} \, \bL ^{c} e^{\wedge \Psi _{c\tilde{c}} [\varphi ] } \wedge \Omega 
+ \pi _{1} \, \bL ^{\tilde{c}} e^{\wedge \Psi _{c\tilde{c}} [\varphi ] } \wedge \widetilde{\Omega } 
+ \pi _{1} \, \bL ^{p} e^{\wedge \Psi _{c\tilde{c}} [\varphi ] } \wedge \Lambda \, . 
\end{align*}

\vspace{2mm} 

Since $\Psi _{c \tilde{c}} [\varphi ]$ satisfies Maurer-Cartan equations ${\bf L}^{c} e^{\wedge \Psi _{c \tilde{c} } [\varphi ]} = 0$ and ${\bf L}^{\tilde{c}} e^{\wedge \Psi _{c \tilde{c} } [\varphi ]} = 0$\,, for any coderivation $\bD$ commuting with ${\bf L}^{c}$ and ${\bf L}^{\tilde{c}}$\,, we find 
\begin{align*}
(-)^{\bD } \bD \, {\bf L}^{c'} e^{\wedge \Psi _{c \tilde{c} } [\varphi ]} = {\bf L}^{c'} e^{\wedge \Psi _{c \tilde{c} } [\varphi ]} \wedge \pi _{1} \bD e^{\wedge \Psi _{c\tilde{c}} [\varphi ] } = 0 \, , \hspace{5mm} ( c' = c\,, \tilde{c} \, ) \,.
\end{align*}
Hence, since $\eta $- and $\tilde{\eta }$-cohomology are trivial, there exist a state $\Psi _{D} [\varphi ]$ such that 
\begin{align*}
- D_{c} D_{\tilde{c}} \Psi _{D} [\varphi ] 
\equiv 
- \pi _{1} {\bf L}^{c} {\bf L}^{\tilde{c}} e^{\wedge \Psi _{c \tilde{c} } [\varphi ]} \wedge
\Psi _{D} [\varphi ] 
= \pi _{1} (-)^{\bD } \bD e^{\wedge \Psi _{c\tilde{c}} [\varphi ] } \, , 
\end{align*}
where we defined $D_{c'} A \equiv - \pi _{1} {\bf L}^{c'} e^{\wedge \Psi _{c \tilde{c} } [\varphi ]} \wedge A
$\,, $(c' = c, \,\tilde{c})$\,, for brevity. 
This is the WZW-like relation for a general $L_{\infty }$ triplet $({\bf L}^{c} , {\bf L}^{\tilde{c} } \,; {\bf L}^{p})$\,, which provides $\delta \Psi _{c\tilde{c}} = - D_{c} D_{\tilde{c} } \Psi _{\delta } $\,, $\partial \Psi _{c\tilde{c}} = - D_{c} D_{\tilde{c} } \Psi _{t} $\,, $\pi _{1} {\bf L}^{p} e^{\wedge \Psi _{c\tilde{c} } } = D_{c} D_{\tilde{c} } \Psi _{L^{p}} $\,, and so on. 
For two coderivations $D_{1}$ and $D_{2}$ which are mutually commute with $\bL ^{c}$ and $\bL ^{\tilde{c}}$\,, we find 
\begin{align*}
\pi _{1} \, \bD _{1} \, \bD _{2} \, e^{\wedge \Psi _{c \tilde{c}} [\varphi ]} 
%& = \pi _{1} \, \bD_{1} \, e^{\wedge \Psi _{c\tilde{c}} [\varphi ]} \wedge \pi _{1} \bD _{2} \big( e^{\wedge \Psi _{c\tilde{c}} [\varphi ] } \big) 
%\no 
& = \pi _{1} \, \bD_{1} \, e^{\wedge \Psi _{c\tilde{c}} [\varphi ]} \wedge \pi _{1} (-)^{\bD _{2}} \bL ^{c} \, \bL ^{\tilde{c}} \, \big( e^{\wedge \Psi _{c\tilde{c}} [\varphi ] } \wedge \Psi _{D_{2}} [\varphi ] \big)  
\no
& = (-)^{\bD _{2} }  \pi _{1} \bL ^{c} \, \bL ^{\tilde{c}} \, \bD _{1} e^{\wedge \Psi _{c\tilde{c}} [\varphi ]} \wedge \Psi _{D_{2}} [\varphi ]  
\no 
& = (-)^{\bD _{2} }  \pi _{1} \bL ^{c} \, \bL ^{\tilde{c}} \Big( e^{\wedge \Psi _{c\tilde{c}} [\varphi ]} \wedge \pi _{1} \bD _{1} \big( e^{\wedge \Psi _{c\tilde{c}} [\varphi ]} \big) \wedge \Psi _{D_{2}} [\varphi ]  \Big)
\no & \hspace{10mm} 
+ (-)^{\bD _{2} }  \pi _{1} \bL ^{c} \, \bL ^{\tilde{c}} \Big( e^{\wedge \Psi _{c\tilde{c}} [\varphi ]} \wedge \pi _{1} \bD _{1} \big( e^{\wedge \Psi _{c\tilde{c}} [\varphi ]} \wedge \Psi _{D_{2}} [\varphi ]  \big) \Big)  \, . 
\end{align*}
It gives general versions of other useful identities derived from the mutual commutativity of coderivations, which are used in the variation of the action. 
For example, $\delta ( \pi _{1} \, {\bf L}^{p} e^{\wedge \Psi _{c \tilde{c}} } ) =  \pi _{1} {\bf L}^{p}(e^{\wedge \Psi _{c \tilde{c}} } \wedge \delta \Psi _{c \tilde{c}} )$\,, (3.2), and (3.9)\,. 
Using these, we find a half of the variation is  
\begin{subequations}
\begin{align}
\label{general variation1}
\la \Psi _{t} , \, \delta \big( \pi _{1} \, \bL ^{p} e^{\wedge \Psi _{c \tilde{c}} } \big) \ra 
& = \la \Psi _{t} , \, \pi _{1} \, \bL ^{p} \big( e^{\wedge \Psi _{c \tilde{c}} } \wedge D_{c} D_{\tilde{c}} \, \Psi _{\delta } \big) \ra 
= - \la \Psi _{\delta } , \, D_{\tilde{c}} D_{c} \pi _{1} \, \bL ^{p} \big( e^{\wedge \Psi _{c \tilde{c}} } \wedge \Psi _{t} \big) \ra 
\no 
& = \la \Psi _{\delta } , \, \partial _{t} \big( \pi _{1} \, \bL ^{p} e^{\wedge \Psi _{c \tilde{c}} } \big) \ra 
+ \la \Psi _{\delta } , \, \pi _{1} \bL ^{c} \bL ^{\tilde{c} } \Big( e^{\wedge \Psi _{c \tilde{c}} } \wedge  D_{c} D_{\tilde{c}} \Psi _{L ^{p}}  \wedge \Psi _{t} \Big) \ra \, . 
\end{align}
We notice that these computation can be carried out by replacing $Q \, \Psi _{\eta \teta } = D_{\eta } D_{\teta } \Psi _{Q}$ of (\ref{2nd term}) with $\pi _{1} \bL ^{p} e^{\wedge \Psi _{c\tilde{c}}} = D_{c} D_{\tilde{c}} \Psi _{L^{p}}$\,. 
Likewise, after short computations, we find 
\begin{align}
\label{general variation2}
\la \delta \Psi _{t} , \, \pi _{1} \, \bL ^{p} e^{\wedge \Psi _{c \tilde{c}} } \ra & = - \la D_{\tilde{c}} D_{c} \delta \Psi _{t} , \, \Psi _{L^{p}} \ra 
\no &
= \la \partial _{t} \big( D_{\tilde{c}} D_{c} \Psi _{\delta } \big) , \, \Psi _{L^{p}} \ra 
+ \la \, \pi _{1} \bL ^{c} \bL ^{\tilde{c} } \Big( e^{\wedge \Psi _{c \tilde{c}} } \wedge  D_{c} D_{\tilde{c}} \Psi _{\delta }  \wedge \Psi _{t} \Big) , \, \Psi _{L^{p}} \ra 
\no &
= - \la \partial _{t} \Psi _{\delta } , \, D_{c} D_{\tilde{c}} \Psi _{L^{p}} \ra 
- \la \Psi _{\delta } , \, \pi _{1} \bL ^{c} \bL ^{\tilde{c} } \Big( e^{\wedge \Psi _{c \tilde{c}} } \wedge  D_{c} D_{\tilde{c}} \Psi _{L ^{p}}  \wedge \Psi _{t} \Big) \ra \, .
\end{align}
\end{subequations} 
Note that this term can be also obtained by replacing $\Psi _{Q}$ of (\ref{1st term}) with $\Psi _{L^{p}}$\,. 
Hence, we obtain the desired result 
\begin{align*}
\delta \la \Psi _{t} [\varphi (t) ] , \, \pi _{1} {\bf L}^{p} e^{\wedge \Psi _{c\tilde{c} } [\varphi (t) ] } \ra  
= (\ref{general variation1}) + (\ref{general variation2})
= \partial _{t} \, \la \Psi _{\delta } [\varphi (t)] , \, \pi _{1} {\bf L}^{p} e^{\wedge \Psi _{c\tilde{c} } [\varphi (t) ] } \ra  \, . 
\end{align*}

We would like to emphasise that the $S_{c\tilde{c}} [\varphi ]$ gives a gauge invariant action for {\it any} $L_{\infty }$ {\it triplet} $( \mathbf{L}^{c} , \mathbf{L}^{\tilde{c}}\, ; \mathbf{L}^{p} )$ in the completely same way. 
In general, field redefinitions $\widehat{\bf U}$ drastically change the string vertices and state space in highly nontrivial manner. 
In terms of $L_{\infty }$ algebras, it is just described by a $L_{\infty }$ morphism between two $L_{\infty }$ triplets, $\widehat{\bf U} : ( \mathbf{L}^{c} , \mathbf{L}^{\tilde{c}}\,; \mathbf{L}^{p} ) \rightarrow ( \mathbf{L}^{c \prime } , \mathbf{L}^{\tilde{c} \prime }\,; \mathbf{L}^{p \prime } )$. 
Hence, the general WZW-like action $S_{c\tilde{c}}[\varphi ]$ is covariant under {\it any string field redefinitions}. 
Thus, as a gauge theory, it may capture general field theoretical properties of superstrings.


\small 

\begin{thebibliography}{99}

  %\cite{Goto:2015pqv}
\bibitem{Goto:2015pqv}
  K.~Goto and H.~Matsunaga,
  ``$A_\infty / L_\infty$ structure and alternative action for WZW-like superstring field theory,''
  JHEP {\bf 1701} (2017) 022
  %doi:10.1007/JHEP01(2017)022
  [arXiv:1512.03379 [hep-th]].
  %%CITATION = doi:10.1007/JHEP01(2017)022;%%
  %5 citations counted in INSPIRE as of 31 Mar 2017

%\cite{Erler:2014eba}
\bibitem{Erler:2014eba}
  T.~Erler, S.~Konopka and I.~Sachs,
  ``NS-NS Sector of Closed Superstring Field Theory,''
  JHEP {\bf 1408} (2014) 158
  %doi:10.1007/JHEP08(2014)158
  [arXiv:1403.0940 [hep-th]].
  %%CITATION = doi:10.1007/JHEP08(2014)158;%%
  %21 citations counted in INSPIRE as of 12 Dec 2016

%\cite{Berkovits:1995ab}
\bibitem{Berkovits:1995ab}
  N.~Berkovits,
  ``SuperPoincare invariant superstring field theory,''
  Nucl.\ Phys.\ B {\bf 450} (1995) 90
   [Erratum-ibid.\ B {\bf 459} (1996) 439]
  [hep-th/9503099].
  %%CITATION = HEP-TH/9503099;%%
  %184 citations counted in INSPIRE as of 26 Jul 2014

%\cite{Erler:2013xta}
\bibitem{Erler:2013xta}
  T.~Erler, S.~Konopka and I.~Sachs,
  ``Resolving Witten`s superstring field theory,''
  JHEP {\bf 1404} (2014) 150
  [arXiv:1312.2948 [hep-th]].
  %%CITATION = ARXIV:1312.2948;%%
  %3 citations counted in INSPIRE as of 26 Jul 2014

%\cite{Berkovits:1998bt}
\bibitem{Berkovits:1998bt}
  N.~Berkovits,
  ``A New approach to superstring field theory,''
  Fortsch.\ Phys.\  {\bf 48} (2000) 31
  [hep-th/9912121].
  %%CITATION = HEP-TH/9912121;%%
  %56 citations counted in INSPIRE as of 26 Jul 2014

%\cite{Okawa:2004ii}
\bibitem{Okawa:2004ii}
  Y.~Okawa and B.~Zwiebach,
  ``Heterotic string field theory,''
  JHEP {\bf 0407} (2004) 042
  [hep-th/0406212].
  %%CITATION = HEP-TH/0406212;%%
  %21 citations counted in INSPIRE as of 26 Jul 2014

%\cite{Berkovits:2004xh}
\bibitem{Berkovits:2004xh}
  N.~Berkovits, Y.~Okawa and B.~Zwiebach,
  ``WZW-like action for heterotic string field theory,''
  JHEP {\bf 0411} (2004) 038
  [hep-th/0409018], %.
  %%CITATION = HEP-TH/0409018;%%
  %27 citations counted in INSPIRE as of 26 Jul 2014

%\cite{Matsunaga:2013mba}
\bibitem{Matsunaga:2013mba}
  H.~Matsunaga,
  ``Construction of a Gauge-Invariant Action for Type II Superstring Field Theory,''
  arXiv:1305.3893 [hep-th]. 
  %%CITATION = ARXIV:1305.3893;%%
  %6 citations counted in INSPIRE as of 26 Jul 2014

%\cite{Matsunaga:2014wpa}
\bibitem{Matsunaga:2014wpa} 
  H.~Matsunaga,
  ``Nonlinear gauge invariance and WZW-like action for NS-NS superstring field theory,''
  JHEP {\bf 1509}, 011 (2015)
  [arXiv:1407.8485 [hep-th]].
  %%CITATION = ARXIV:1407.8485;%%
  %7 citations counted in INSPIRE as of 26 Oct 2015

%\cite{Sen:2015uaa}
\bibitem{Sen:2015uaa} 
  A.~Sen,
  ``BV Master Action for Heterotic and Type II String Field Theories,''
  arXiv:1508.05387 [hep-th].
  %%CITATION = ARXIV:1508.05387;%%
  %2 citations counted in INSPIRE as of 26 Oct 2015

%%\cite{Saroja:1992vw}
\bibitem{Saroja:1992vw}
  R.~Saroja and A.~Sen,
  ``Picture changing operators in closed fermionic string field theory,''
  Phys.\ Lett.\ B {\bf 286} (1992) 256
  [hep-th/9202087].
%  %%CITATION = HEP-TH/9202087;%%
%  %14 citations counted in INSPIRE as of 26 Jul 2014

%%\cite{Jurco:2013qra}
\bibitem{Jurco:2013qra}
  B.~Jurco and K.~Muenster,
  ``Type II Superstring Field Theory: Geometric Approach and Operadic Description,''
  JHEP {\bf 1304} (2013) 126
  [arXiv:1303.2323 [hep-th]].
%  %%CITATION = ARXIV:1303.2323;%%
%  %7 citations counted in INSPIRE as of 29 Jul 2014

%%\cite{Witten:1986qs}
\bibitem{Witten:1986qs}
  E.~Witten,
  ``Interacting Field Theory of Open Superstrings,''
  Nucl.\ Phys.\ B {\bf 276} (1986) 291.
%  %%CITATION = NUPHA,B276,291;%%
%  %449 citations counted in INSPIRE as of 29 Jul 2014

%%\cite{Wendt:1987zh}
\bibitem{Wendt:1987zh}
  C.~Wendt,
  ``Scattering Amplitudes and Contact Interactions in Witten's Superstring Field Theory,''
  Nucl.\ Phys.\ B {\bf 314} (1989) 209.
%  %%CITATION = NUPHA,B314,209;%%
%  %83 citations counted in INSPIRE as of 29 Jul 2014

%\cite{Zwiebach:1992ie}
\bibitem{Zwiebach:1992ie}
  B.~Zwiebach,
  ``Closed string field theory: Quantum action and the B-V master equation,''
  Nucl.\ Phys.\  B {\bf 390}, 33 (1993)
  [arXiv:hep-th/9206084].

%\cite{Witten:1985cc}
%\bibitem{Witten:1985cc}
%  E.~Witten,
%  ``Noncommutative Geometry and String Field Theory,''
%  Nucl.\ Phys.\  B {\bf 268}, 253 (1986).
  %%CITATION = NUPHA,B268,253;%%

%%\cite{Schubert:1991en}
\bibitem{Schubert:1991en} 
  C.~Schubert,
  ``The Finite gauge transformations in closed string field theory,''  Lett.\ Math.\ Phys.\  {\bf 26}, 259 (1992).
%  %%CITATION = LMPHD,26,259;%%

%\cite{Kunitomo:2015usa}
\bibitem{Kunitomo:2015usa} 
  H.~Kunitomo and Y.~Okawa,
  ``Complete action of open superstring field theory,''
  arXiv:1508.00366 [hep-th].
  %%CITATION = ARXIV:1508.00366;%%
  %3 citations counted in INSPIRE as of 26 Oct 2015

\bibitem{GK}
  K.~Goto and H.~Kunitomo,
  ``Construction of action for heterotic string field theory including the Ramond sector,''
  arXiv:1606.07194 [hep-th].
  %%CITATION = ARXIV:1606.07194;%%

%\cite{Matsunaga:2015kra}
\bibitem{Matsunaga:2015kra}
  H.~Matsunaga,
  ``Comments on complete actions for open superstring field theory,''
  JHEP {\bf 1611} (2016) 115
  %doi:10.1007/JHEP11(2016)115
  [arXiv:1510.06023 [hep-th]].
  %%CITATION = doi:10.1007/JHEP11(2016)115;%%
  %5 citations counted in INSPIRE as of 08 Dec 2016

%\cite{Erler:2015rra}
\bibitem{Erler:2015rra} 
  T.~Erler, Y.~Okawa and T.~Takezaki,
  ``$A_\infty$ structure from the Berkovits formulation of open superstring field theory,''
  arXiv:1505.01659 [hep-th].
  %%CITATION = ARXIV:1505.01659;%%

%\cite{Erler:2015uba}
\bibitem{Erler:2015uba} 
  T.~Erler,
  ``Relating Berkovits and $A_\infty$ Superstring Field Theories; Small Hilbert Space Perspective,''
  arXiv:1505.02069 [hep-th].
  %%CITATION = ARXIV:1505.02069;%%

%\cite{Erler:2015uoa}
\bibitem{Erler:2015uoa} 
  T.~Erler,
  ``Relating Berkovits and $A_\infty$ Superstring Field Theories; Large Hilbert Space Perspective,''
  arXiv:1510.00364 [hep-th].
  %%CITATION = ARXIV:1510.00364;%%
  %1 citations counted in INSPIRE as of 26 Oct 2015

%\cite{Goto:2015hpa}
\bibitem{Goto:2015hpa}
  K.~Goto and H.~Matsunaga,
  ``On-shell equivalence of two formulations for superstring field theory,''
  arXiv:1506.06657 [hep-th].
  %%CITATION = ARXIV:1506.06657;%%
  %5 citations counted in INSPIRE as of 26 Oct 2015

\bibitem{Erler}
  T.~Erler, 
  in preparation. 


\end{thebibliography}
\end{document}